\documentstyle[epsf,rotate,12pt]{article}

\begin{document}

\setlength{\baselineskip}{3ex}
\bibliographystyle{unsrt}
\thispagestyle{empty}
\frenchspacing

\newcommand{\Bbb}{\bf}
\newcommand{\QCD}{$\mbox{QCD}_{3+1}$}
\newcommand{\QED}{$\mbox{QED}_{2+1}$}
\newcommand{\gtrsim}{\displaystyle\mathop{>}_{\sim}}
\newcommand{\lesssim}{\displaystyle\mathop{<}_{\sim}}
\newcommand{\ERI}{effective residual interaction}

\hyphenation{ei-gen-value}

\begin{center}


{\Large {\bf
Exploring the $\pi^+$--$\pi^+$ Interaction in Lattice QCD
}} \vspace{4ex}

{\normalsize

\noindent


H.R. Fiebig$^{\mbox{\scriptsize a}}$,
K. Rabitsch$^{\mbox{\scriptsize b}}$,
H. Markum$^{\mbox{\scriptsize b}}$,
A. Mih\'{a}ly$^{\mbox{\scriptsize c}}$

\vspace{3ex}


\noindent ${}^{\mbox{\scriptsize a}}$
Physics Department, FIU--University Park,
Miami, Florida 33199, USA

\noindent ${}^{\mbox{\scriptsize b}}$
Institut f{\"u}r Kernphysik, Technische Universit{\"a}t Wien,
A-1040 Vienna, Austria

\noindent ${}^{\mbox{\scriptsize c}}$
Department of Theoretical Physics, Lajos Kossuth University,
H-4010 Debrecen, Hungary


\vspace{3ex} (Revised 11-Nov-1999) } \vspace{3ex}


{\bf Abstract} \vspace{-1ex} \end{center}

An \ERI\ for a meson-meson system is computed in lattice QCD. We describe
the theoretical framework and present its application to the $I=2$ channel
S-wave interaction of the $\pi$--$\pi$ system. Scattering phase shifts
are also computed and compared to experimental results.

\vspace*{\fill}


\section{Introduction}

For many decades boson exchange models
have served as a theoretical basis for modeling the 
strong interaction \cite{Yuk35}. This class of models has reached
a high level of sophistication and has become a standard tool encompassing
a large variety of strong
interaction phenomena at low and intermediate energies \cite{Mac89}.
On the other hand it has also become increasingly clear that
quantum chromodynamics (QCD) provides the microscopic underpinnings for 
hadronic interaction.

The strong nuclear force operates in an energy 
range where the intrinsic hadronic ground states and a few
excited states are important. At this low-to-intermediate energy scale
the physics of QCD is only accessible through nonperturbative methods.
While quark models have been used to capture part of the
microscopic physics \cite{Fae83}, there is no substitute for exploring 
hadronic interaction on a fundamental level.
Currently the most appropriate way to do this is lattice
field theory, supplemented by computational techniques.

As of today lattice QCD is well developed and explores many different
topics ranging from quark confinement to hadron masses.
Hadronic interactions, however,
still have not moved into the mainstream of activity. Early work of 
Richards and collaborators \cite{Ric89,Ric90} addresses the role of
quark exchange diagrams for meson-meson systems. L\"{u}scher's formula
\cite{Lue86}, which exploits finite-size effects,
allows the extraction of scattering lengths from the lattice.
It was applied to systems of pions and nucleons \cite{Fuk95} within
lattice QCD. Beyond static properties of two-hadron systems
theoretical concepts become more involved.
L\"{u}scher has devised a method \cite{Lue91a} for relating the excitation
spectrum of a two-body system in a finite box to scattering phase shifts.
Elastic channel resonances are also within its scope \cite{Lue91b}.
Phase shifts become available for a discrete set of momenta.
Discrete interpolation between those can be done through the use of
different-sized lattices.
Within an $O(4)$ symmetric $\phi^4$ model extensive numerical tests
\cite{Goe94} have shown that the method works well.
However, unlike with the previous model, stringent numerical standards
are much harder to meet for applications with QCD, where
gauge fields and composite asymptotic states complicate the issue.
In L\"{u}scher's method the `master' equation that relates the
two-body energy spectrum to scattering phases, unfortunately,
has multiple solutions,
the physical ones can be difficult to discern by solely numerical means.

In the nonrelativistic limit interactions can be described by potentials.
Ongoing lattice work by Green and coworkers \cite{Gre98} focuses on geometric 
configurations of quarks in hadron-hadron systems and their interaction 
energies. A rather large body of work exists
\cite{Gre98}\nocite{Pen99a,Pen99,Gre95,Gre93b}-\cite{Gre93a}
with $SU(2)$ as the gauge group, including the heavy-light meson-meson
system studied by Steward and Koniuk \cite{Ste98}.

In $SU(3)$ a simple geometric model with static quarks tried by Rabitsch
et al. \cite{Rab93} provides some exploratory insight into the gluon-generated
part of the interaction between two three-quark clusters.
Extending this line of work, studies of heavy-light meson-meson
systems were done \cite{Mih97} where the heavy quarks are treated as static 
and the light valence quarks are dynamic in the sense that their propagation
is computed from the fermionic, staggered, lattice action. Although these
studies are numerically difficult it appears that an attractive force
between the partners is a common feature. However, in line with the
exploratory flavor of these studies, the assignment of real-world
quantum numbers to the hadronic systems was neglected. 

Two-hadron systems with one infinitely heavy quark in each hadron have the advantage
that their relative distance $r$ is a theoretically well-defined notion.
This opens the door to interactions of heavy-light systems. Still, realistic cases,
involving B mesons for example, do require massive computing power. Most
recently the UKQCD group has taken up simulations in this direction \cite{Mic99}.

For light-light systems some work has been done in dimensions smaller
than $d+1=4$ and gauge groups simpler than $SU(3)$.
Probably the simplest nontrivial
system in this class is a planar, $d=2$ lattice theory with a $U(1)$ gauge
group. In finite volume, and with properly chosen coupling, $\beta$, the
theory is confining. It can be used as a `toy' system to develop
techniques for studying hadronic interaction. This was done within the
staggered fermion scheme for a meson-meson system \cite{Can97}. 
In the nonrelativistic approximation a potential can be computed and
used in a Schr{\"o}dinger equation to calculate scattering phase shifts.

We will here follow essentially the `blueprint' of \cite{Can97} applying it
to a $\pi$--$\pi$ system in the isospin $I=2$ channel. Using an $SU(3)$ gauge
group in $d=3$ dimensions and a highly improved action we attempt a realistic
calculation of the residual $\pi$--$\pi$ long-to-intermediate range interaction.
Extrapolation to the chiral limit and a comparison 
of the corresponding scattering phase shifts to experimental results are performed.
 
Preliminary stages of this work have been reported before \cite{Fie99a,Rab97}.
A self-contained presentation of the formalism,
in section~\ref{secForm}, and
a discussion and interpretation of the lattice results, in 
section~\ref{secComp}, are the purpose of this paper.

\section{Formalism}\label{secForm}

In the context of scattering within the framework of a quantum field
theory the LSZ reduction formalism comes to mind. Putting the concept of `in'
and `out' states, as $t\rightarrow \pm\infty$ in Minkowski time, to
use in the Euclidean formulation proves, unfortunately, prone to theoretical
difficulties. Maiani and Testa \cite{Mai90} studied a 3-point function
$\langle \phi_{q_1}(t_1) \phi_{q_2}(t_2) J(0) \rangle $
involving a local source $J$ and two fields $\phi$ interpolating a
pseudoscalar particle. They showed that in the region $0\ll t_2 \ll t_1 $,
of Euclidean times, the 3-point function depends only on the
average (sum) of `in' and `out' matrix elements. Relative phases containing
physical scattering information can thus not be extracted.
This discouraging result, sometimes referred to as a no-go theorem, does not
mean, however, that it is in principle impossible to extract physical
two-body information from the lattice.
The no-go theorem can be avoided in special cases \cite{Cui96}, and also
bypassed entirely by using a different method. A case in point is the
aforementioned formalism by L\"{u}scher \cite{Lue91a}. It is based on a standard
mass calculation involving the two-body system in a finite box, say with
periodic boundary conditions. The discrete
two-body energy spectrum obtained from a lattice simulation 
comprises the residual interaction provided the box is
large enough so that the particles (hadrons) can separate from each other.
In L\"{u}scher's original approach the spectrum of the interacting
two-body system is in some sense
`compared' to that of free particles in the same box, their propagation
being represented by
`singular periodic solutions' of a Helmholtz equation. The method employed in
the present work is similar in spirit to L\"{u}scher's to the extent that it is based on
a comparison of two-body excitation spectra in a finite volume.
We will, however, analyze the two-body information from the lattice in different ways.
First, instead of using `singular periodic solutions' to represent the
noninteracting system, we construct a two-body system of
noninteracting {\em composite} mesons within our lattice simulation.
Second, we aim at a (nonrelativistic) potential between mesons 
as a means of interpolation between the lattice sites.

In this section we describe the strategy for computing an \ERI\ 
between composite hadrons from the lattice.
The formalism will be set up for the example of two pseudoscalar pions,
however, extensions to other hadron-hadron systems should be straightforward:
A set of interpolating fields for the two-hadron system is used to
compute a time correlation matrix for the full (interacting) system.
It has a time correlation matrix for the free (noninteracting) system
as an additive term. Comparison between those two correlators contains
information about the effective residual interaction\footnote{We closely
follow \protect\cite{Can97} in this section.}.

\subsection{Meson-Meson Fields}

Starting with the one-meson field we choose
\begin{equation}
\phi_{\vec{p}}(t)=L^{-3}\sum_{\vec{x}}\,e^{i\vec{p}\cdot\vec{x}}\,
\bar{\psi}^{\{ S\}}_{dA}(\vec{x},t)\,\gamma_5\,\psi^{\{ S\}}_{uA}(\vec{x},t) \,,
\label{onephi}\end{equation}
where $L^3$ is the space lattice volume, and $\psi^{\{ S\}}_{fA\mu}(\vec{x},t)$
is a linear combination of quark fields
${\psi}_{fA\mu}(\vec{x},t)$ with flavor $f$, color $A$, and Dirac index $\mu$,
at lattice site $x=(\vec{x},t)$. The field is local, however,
the superscript ${}^{\{S\}}$ indicates that the field is `smeared', iterated
$S$ times. The purpose of smearing \cite{Ale94} is to enhance the amplitude
for propagation of the hadron in the ground state.

Products of those fields probe for multi-meson excitations. Our choice is
\begin{equation}
\Phi_{\vec{p}}(t)=\phi_{-\vec{p}}(t)\,\phi_{+\vec{p}}(t) \,.
\label{twophi}\end{equation}
These operators describe two-meson systems with total momentum 
$\vec{P}=0$ and relative momenta $\vec{p}$. They are designed to excite
degrees of freedom of relative motion of the two composite mesons, in their
respective intrinsic ground states. On a periodic lattice the momenta are
\begin{equation}
\vec{p}=\frac{2\pi}{L}\vec{k} \quad\mbox{with}\quad \vec{k}\in{\Bbb R}^3 \,.
\label{momenta}\end{equation}

\subsection{Correlation Matrices}

Correlation matrices describe the propagation of the above fields
in Euclidean time. Since we desire a comparison of
the free (noninteracting) and the full (interacting) systems two types
of correlation matrices are needed.

The time-correlation matrix for the one-meson system (2-point function) is 
\begin{equation} 
C^{(2)}_{\vec{p}\,\vec{q}}(t,t_0)=
\langle\phi^{\dagger}_{\vec{p}}(t)\,\phi_{\vec{q}}(t_0)\rangle-
\langle\phi^{\dagger}_{\vec{p}}(t)\rangle
\langle\phi^{\phantom{\dagger}}_{\vec{q}}(t_0)\rangle \,,
\label{eq41}\end{equation}
with $\langle\;\rangle$ denoting the gauge-configuration average.
In the usual way it is readily worked out by means of Wick's theorem
\begin{equation} 
\ldots \stackrel{n\rule{5mm}{0mm}}{\psi^{\{S\}}}(x)
\stackrel{n\rule{5mm}{0mm}}{\bar{\psi}^{\{S\}}}(y)\ldots =
\ldots \stackrel{n\rule{5mm}{0mm}}{G^{\{S\}}}(x,y)\ldots \,,
\label{contrac}\end{equation}
where the pair $n\,n$ on the l.h.s. defines the partners of a 
contraction. We assume the quark propagator $G^{\{S\}}$ of the smeared fields
to be flavor-independent
\begin{equation}
G^{\{S\}}_{fA\mu,gB\nu}(x,y)=\delta_{fg}\,G^{\{S\}}_{A\mu,B\nu}(x,y) \,.
\label{Gflavor}\end{equation}
Since the smearing prescription used \cite{Ale94} does not mix flavor
indices, (\ref{Gflavor}) is a consequence of the corresponding relation
of the unsmeared fields and thus holds independently of $S$, see
Appendix \ref{appRanSrc}.
We therefore will omit ${}^{\{S\}}$ from $G$ to simplify notation.
In (\ref{eq41}) the separable term is zero, while there appear two
contractions, $n=1,2$. Making use of translational invariance
one obtains
\begin{equation}
C^{(2)}_{\vec{p}\,\vec{q}}(t,t_0)=
\delta_{\vec{p}\,\vec{q}}\;c_{\vec{p}}(t,t_0)\,e^{i\vec{p}\cdot\vec{x}_0} \,,
\label{eq58}\end{equation}
with
\begin{equation}
c_{\vec{p}}(t,t_0)=
L^{-3}\sum_{\vec{x}}e^{-i\vec{p}\cdot\vec{x}}
\langle|G_{A\mu,B\nu}(\vec{x}t,\vec{x}_0t_0)|^2\rangle \,.
\label{eq44}\end{equation}
Sums over repeated indices are understood.
The source point $x_0=(\vec{x}_0t_0)$ is fixed and arbitrary.
For later purposes we note that
\begin{equation}
c_{-\vec{p}}(t,t_0)=c^{\ast}_{\vec{p}}(t,t_0) \,.
\label{eq59}\end{equation}

\begin{figure}[htb]
\begin{center}
\mbox{\epsfxsize=80mm\epsfbox{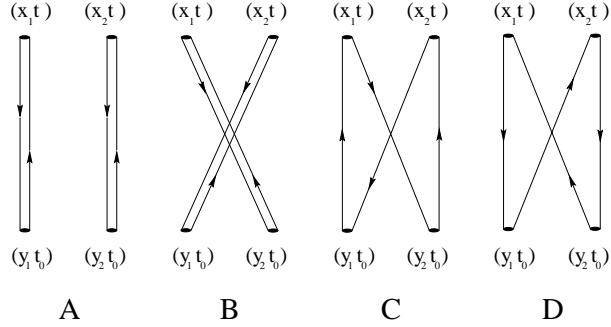}}
\end{center}
\caption{Diagrammatic classification of the full correlation matrix
$C^{(4)}(t,t_0)$ according to (\protect\ref{eq46})-(\protect\ref{eq47}).}
\label{figABCD}\end{figure}

The full meson-meson system propagates according to the 4-point function 
\begin{equation}
C^{(4)}_{\vec{p}\,\vec{q}}(t,t_0)=
\langle\Phi^{\dagger}_{\vec{p}}(t)\,\Phi_{\vec{q}}(t_0)\rangle-
\langle\Phi^{\dagger}_{\vec{p}}(t)\rangle
\langle\Phi^{\phantom{\dagger}}_{\vec{q}}(t_0)\rangle \,.
\label{eq45}\end{equation}
Working out the contractions here leads to a more complicated expression
\begin{eqnarray}
C^{(4)}_{\vec{p}\,\vec{q}} (t,t_0) &=& L^{-12} \: \sum_{\vec x_1} \: 
\sum_{\vec x_2} \: \sum_{\vec y_1} \: \sum_{\vec y_2} \: {\rm e}^{i\vec p 
\cdot(\vec x_2 -
\vec x_1) + i \vec q \cdot (\vec y_2 - \vec y_1)}
\rule{3.0cm}{0mm} \nonumber \\
& & \mbox{} \left\langle \phantom{x}
G^\ast_{A_2\mu_2,B_2\nu_2}(\vec x_2t, \vec y_2t_0) \quad
G_{A_2\mu_2,B_2\nu_2}(\vec x_2t, \vec y_2t_0) \quad \times
\right. \nonumber \\
& & \rule{1.5cm}{0mm}
G^\ast_{A_1\mu_1,B_1\nu_1}(\vec x_1t, \vec y_1t_0) \quad
G_{A_1\mu_1,B_1\nu_1}(\vec x_1t, \vec y_1t_0) \nonumber \\
& & \mbox{} + 
G^\ast_{A_1\mu_1,B_2\nu_2}(\vec x_1t, \vec y_2t_0) \quad
G_{A_1\mu_1,B_2\nu_2}(\vec x_1t, \vec y_2t_0) \quad \times \nonumber \\
& & \rule{1.5cm}{0mm}
G^\ast_{A_2\mu_2,B_1\nu_1}(\vec x_2t, \vec y_1t_0) \quad
G_{A_2\mu_2,B_1\nu_1}(\vec x_2t, \vec y_1t_0) \nonumber \\
& & \mbox{} - 
G^\ast_{A_2\mu_2,B_1\nu_1}(\vec x_2t, \vec y_1 t_0) \quad
G_{A_2\mu_2,B_2\nu_2}(\vec x_2t, \vec y_2 t_0) \quad \times \nonumber \\
& & \rule{1.5cm}{0mm}
G^\ast_{A_1\mu_1,B_2\nu_2}(\vec x_1t, \vec y_2 t_0) \quad
G_{A_1\mu_1,B_1\nu_1}(\vec x_1t, \vec y_1 t_0) \nonumber \\
& & \mbox{} - 
G_{A_2\mu_2,B_1\nu_1}(\vec x_2t, \vec y_1 t_0) \quad
G^\ast_{A_2\mu_2,B_2\nu_2}(\vec x_2t, \vec y_2 t_0) \quad \times \nonumber \\
& & \rule{1.5cm}{0mm} \left.
G_{A_1\mu_1,B_2\nu_2}(\vec x_1t, \vec y_2 t_0) \quad
G^\ast_{A_1\mu_1,B_1\nu_1}(\vec x_1t, \vec y_1 t_0)
\phantom{x} \right\rangle \,. \label{eq46}
\end{eqnarray}
Again, the separable term in (\ref{eq45}) vanishes. Figure~\ref{figABCD}
shows the diagrammatic classification of (\ref{eq46}),
leading to four terms
\begin{equation}
C^{(4)} = C^{(4A)} + C^{(4B)} - C^{(4C)} - C^{(4D)} \,.
\label{eq47}\end{equation}

It is useful to trace the pattern of contractions back to the `meson content'
of the correlators. Denoting the contractions as in (\ref{contrac}) we write
\begin{equation}
C^{(4A)}=\langle \stackrel{43\phantom{9}}{\phi^{\dagger}_{+\vec{p}}}
            \stackrel{21\phantom{99}}{\phi^{\dagger}_{-\vec{p}}}
        \stackrel{12\phantom{99}}{\phi^{\phantom{\dagger}}_{-\vec{q}}}
    \stackrel{34\phantom{9}}{\phi^{\phantom{\dagger}}_{+\vec{q}}} \rangle
        =\langle \stackrel{43\phantom{9}}{\phi^{\dagger}_{+\vec{p}}}
             \stackrel{34\phantom{9}}{\phi^{\phantom{\dagger}}_{+\vec{q}}}
        \stackrel{21\phantom{99}}{\phi^{\dagger}_{-\vec{p}}}
    \stackrel{12\phantom{99}}{\phi^{\phantom{\dagger}}_{-\vec{q}}} \rangle
\label{eq49}\end{equation}
\begin{equation} 
C^{(4B)}=\langle \stackrel{21\phantom{9}}{\phi^{\dagger}_{+\vec{p}}}
            \stackrel{43\phantom{99}}{\phi^{\dagger}_{-\vec{p}}}
        \stackrel{12\phantom{99}}{\phi^{\phantom{\dagger}}_{-\vec{q}}}
    \stackrel{34\phantom{9}}{\phi^{\phantom{\dagger}}_{+\vec{q}}} \rangle
        =\langle \stackrel{43\phantom{99}}{\phi^{\dagger}_{-\vec{p}}}
             \stackrel{34\phantom{9}}{\phi^{\phantom{\dagger}}_{+\vec{q}}}
         \stackrel{21\phantom{9}}{\phi^{\dagger}_{+\vec{p}}}
     \stackrel{12\phantom{99}}{\phi^{\phantom{\dagger}}_{-\vec{q}}} \rangle \,.
\label{eq410}\end{equation}
Each pair of equal numbers ${n}={1}\ldots {4}$ identifies contracted quark
fields as they emerge from (\ref{onephi}).
Diagrams C and D exhibit valence quark exchange between the mesons 
\begin{equation} 
C^{(4C)}=\langle \stackrel{23\phantom{9}}{\phi^{\dagger}_{+\vec{p}}}
             \stackrel{41\phantom{99}}{\phi^{\dagger}_{-\vec{p}}}
         \stackrel{12\phantom{99}}{\phi^{\phantom{\dagger}}_{-\vec{q}}}
     \stackrel{34\phantom{9}}{\phi^{\phantom{\dagger}}_{+\vec{q}}} \rangle
\label{eq411}\end{equation}
\begin{equation} 
C^{(4D)}=\langle \stackrel{41\phantom{9}}{\phi^{\dagger}_{+\vec{p}}}
             \stackrel{23\phantom{99}}{\phi^{\dagger}_{-\vec{p}}}
         \stackrel{12\phantom{99}}{\phi^{\phantom{\dagger}}_{-\vec{q}}}
     \stackrel{34\phantom{9}}{\phi^{\phantom{\dagger}}_{+\vec{q}}} \rangle \,,
\label{eq412}\end{equation}
and thus must be considered as sources of effective residual interaction.

We wish to extract from the full correlator $C^{(4)}$ a part 
which describes
two composite `lattice' mesons with their residual interaction `switched off'.
This part is entirely contained in diagrams A and B, but
gluonic correlations between the mesons are
also present because the gauge configuration average $\langle\;\rangle$
is taken over the product of all four fields.
Gluonic correlations contribute to the \ERI.
It is thus necessary to isolate the uncorrelated
part contained in $C^{(4A)}+C^{(4B)}$.

\subsection{Free Meson-Meson Correlator}

\begin{figure}[htb]
\begin{center}
\mbox{\epsfxsize=130mm\epsfbox{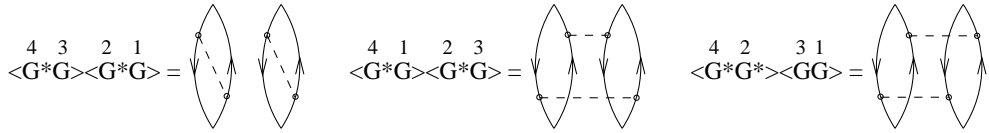}}
\end{center}
\caption{Illustration of the cumulant expansion of $C^{(4A)}$. The solid
lines represent quark propagators, and the dashed lines gluonic correlations.}
\label{figCUMUL}\end{figure}

Towards this end consider $C^{(4A)}$ in the form of (\ref{eq49}). In the spirit
of (\ref{contrac}) we have
\begin{equation} 
C^{(4A)} \sim \langle
\stackrel{4}{G^{\ast}} \stackrel{3}{G^{\phantom{\ast}}}\!
\stackrel{2}{G^{\ast}} \stackrel{1}{G^{\phantom{\ast}}}\! \rangle \,,
\label{eq51}\end{equation}
where the $\sim$ indicates the Fourier sums, etc., which carry over from
(\ref{onephi}), see (\ref{eq46}). 
The gauge configuration average in (\ref{eq51}) may be analyzed 
in a systematic manner by means of cumulant expansion \cite{MaS85}.
Taking advantage of $\langle G\rangle=0$ we have
\begin{equation} 
\langle \stackrel{4}{G^{\ast}} \stackrel{3}{G^{\phantom{\ast}}}
\stackrel{2}{G^{\ast}} \stackrel{1}{G^{\phantom{\ast}}} \rangle = 
\langle \stackrel{4}{G^{\ast}} \stackrel{3}{G^{\phantom{\ast}}} \rangle 
\langle \stackrel{2}{G^{\ast}} \stackrel{1}{G^{\phantom{\ast}}} \rangle +
\langle \stackrel{4}{G^{\ast}} \stackrel{1}{G^{\phantom{\ast}}} \rangle 
\langle \stackrel{2}{G^{\ast}} \stackrel{3}{G^{\phantom{\ast}}} \rangle +
\langle \stackrel{4}{G^{\ast}} \stackrel{2}{G^{\ast}} 
\rangle \langle  \stackrel{3}{G^{\phantom{\ast}}}
\stackrel{1}{G^{\phantom{\ast}}} \rangle +
\langle\!\langle \stackrel{4}{G^{\ast}} \stackrel{3}{G^{\phantom{\ast}}}
\stackrel{2}{G^{\ast}} \stackrel{1}{G^{\phantom{\ast}}}\!\!\rangle\!\rangle \,.
\label{eq52}\end{equation}
The last term defines the cumulant.
The first three terms on the right-hand side of (\ref{eq52})
are illustrated in Fig.~\ref{figCUMUL}.
 
The dashed lines indicate that the quark propagators are correlated
through gluons.
Evidently only the first one of the three separable terms in (\ref{eq52})
represents free, uncorrelated, mesons.
All other terms are sources of \ERI\ 
between the mesons.  We therefore define
\begin{equation} 
\overline{C}^{(4A)}=
\langle \stackrel{43\phantom{9}}{\phi^{\dagger}_{+\vec{p}}}
        \stackrel{34\phantom{9}}{\phi^{\phantom{\dagger}}_{+\vec{q}}} \rangle
\langle \stackrel{21\phantom{99}}{\phi^{\dagger}_{-\vec{p}}}
        \stackrel{12\phantom{99}}{\phi^{\phantom{\dagger}}_{-\vec{q}}} \rangle
\sim
\langle \stackrel{4}{G^{\ast}} \stackrel{3}{G^{\phantom{\ast}}} \rangle 
\langle \stackrel{2}{G^{\ast}} \stackrel{1}{G^{\phantom{\ast}}} \rangle \,.
\label{eq53}\end{equation}
A similar analysis of $C^{(4B)}$ leads to
\begin{equation} 
\overline{C}^{(4B)}=
\langle \stackrel{43\phantom{99}}{\phi^{\dagger}_{-\vec{p}}}
        \stackrel{34\phantom{9}}{\phi^{\phantom{\dagger}}_{+\vec{q}}} \rangle
\langle \stackrel{21\phantom{9}}{\phi^{\dagger}_{+\vec{p}}}
       \stackrel{12\phantom{99}}{\phi^{\phantom{\dagger}}_{-\vec{q}}}\rangle\,.
\phantom{ \sim
\langle \stackrel{4}{G^{\ast}} \stackrel{3}{G^{\phantom{\ast}}} \rangle 
\langle \stackrel{2}{G^{\ast}} \stackrel{1}{G^{\phantom{\ast}}} \rangle }
\label{eq54}\end{equation}
The sum of those is the free meson-meson time correlation matrix
\begin{equation} 
\overline{C}^{(4)}_{\vec{p}\,\vec{q}}(t,t_0) =
\overline{C}^{(4A)}_{\vec{p}\,\vec{q}}(t,t_0) +
\overline{C}^{(4B)}_{\vec{p}\,\vec{q}}(t,t_0) \,,
\label{eq55}\end{equation}
which is an additive part of the full 4-point correlator
\begin{equation} 
C^{(4)}=\overline{C}^{(4)}+C^{(4)}_I \,.
\label{eq56}\end{equation}
The remainder $C^{(4)}_I$, defined through (\ref{eq56}),
comprises all sources of the residual interaction, 
be it from gluonic correlations, or quark exchange (or quark-antiquark loops,
if the simulation is unquenched).

The free correlator $\overline{C}^{(4)}$ describes two noninteracting 
identical lattice mesons. Their intrinsic
structure is consistent with the dynamics 
determined by the lattice field model and its numerical implementation.
Naturally, $\overline{C}^{(4)}$ may be expressed in terms of the 2-point
correlator $C^{(2)}$. Using (\ref{eq41}) and (\ref{eq53}),(\ref{eq54}) gives
\begin{equation} 
\overline{C}^{(4)}_{\vec{p}\,\vec{q}}=
C^{(2)}_{\vec{p},\vec{q}}\,C^{(2)}_{-\vec{p},-\vec{q}}+ 
C^{(2)}_{-\vec{p},\vec{q}}\,C^{(2)}_{\vec{p},-\vec{q}} \,. 
\label{eq57}\end{equation}
Continuing with (\ref{eq58}) and (\ref{eq59}) this becomes
\begin{equation} 
\overline{C}^{(4)}_{\vec{p}\,\vec{q}}(t,t_0)=
\left( \delta_{\vec{p},\vec{q}}+\delta_{-\vec{p},\vec{q}} \right)
\left| c_{\vec{p}}(t,t_0) \right|^2 \,.
\label{eq510}\end{equation}
It is interesting to note that the property
\begin{equation} 
\overline{C}^{(4)}_{\vec{p},\vec{q}}=\overline{C}^{(4)}_{-\vec{p},\vec{q}}=
\overline{C}^{(4)}_{\vec{p},-\vec{q}}=\overline{C}^{(4)}_{-\vec{p},-\vec{q}} \,,
\label{eq511}\end{equation}
which is evident from (\ref{eq57}), and also holds for $C^{(4)}$ through
inspection of (\ref{eq46}), reflects Bose symmetry with respect to 
the {\em composite} mesons. Permutation of the mesons, one with momentum 
$+\vec{p}$ the other with momentum $-\vec{p}$, results in the substitution 
$\vec{p}\rightarrow -\vec{p}$.

\subsection{Residual Interaction}

Eventually it would be desirable to extract scattering amplitudes,
t-matrix elements, directly from a numerical simulation of lattice QCD.
This would require a formulation of the LSZ prescription appropriate for
the Euclidean lattice with the added complication that the asymptotic states
are composites which themselves are determined by the discretized
quantum field theory and its specific numerical implementation.
We are presently not aware of such a formulation.

For the time being we are content with defining an effective interaction
from the computed lattice correlation matrices $C^{(4)}$ and
$\overline{C}^{(4)}$ along the lines of Appendix \ref{appElem}.
We have in mind a description of the meson-meson
system in terms of an effective Hamiltonian 
\begin{equation}
{\cal H}={\cal H}_0+{\cal H}_I \,,
\label{Heff}\end{equation}
where ${\cal H}_0$ describes free propagation and ${\cal H}_I$ the
residual interaction.
We here adopt a definition of ${\cal H}_I$ suggested by
an elementary Bose field theory subject to canonical
quantization. In such a theory it is straightforward 
to calculate the perturbative expansion of $C^{(4)}$, as defined in
(\ref{eq45}) together with (\ref{twophi}), but using elementary Bose
fields instead. This is outlined in Appendix \ref{appElem}.
The zero-order term of the expansion corresponds
to $\overline{C}^{(4)}$. Working up to second order we are lead to the
following strategy:

Define an effective correlation matrix
\begin{eqnarray} 
{\cal C}(t,t_0)
&=& {\overline{C}^{(4)}(t,t_0)}^{-1/2}\,
{C}^{(4)}(t,t_0)\,\,{\overline{C}^{(4)}(t,t_0)}^{-1/2}
\label{Ceff1}
\\ &=& {\overline{C}^{(4)}(t,t_0)}^{-1/2}\, 
{C}_I^{(4)}(t,t_0)\,\,{\overline{C}^{(4)}(t,t_0)}^{-1/2} + {Bbb 1} \,,
\label{Ceff2}\end{eqnarray}
where (\ref{eq56}) was used for the alternative
form.\footnote{Dividing out square roots of correlators $\overline{C}^{(4)}$ for
composite free mesons bears some similarity to
`amputating' dressed external propagator lines.}
Then in terms of ${\cal C}$ we define
\begin{equation} 
{\cal H}_I = - \lim_{t\rightarrow\infty} \frac{\partial
\ln{\cal C}(t,t_0)}{\partial t} \,.
\label{ERI}\end{equation}
To obtain ${\cal H}_I$ numerical diagonalization of the
effective correlator
matrix ${\cal C}$ is called for. The eigenvalues of ${\cal C}$ are expected
to behave exponentially with $t$, thus rendering ${\cal H}_I$ a
time-independent matrix. 

In the numerical simulation it is desirable to utilize time slices with $t$ as
`early' as possible in (\ref{ERI}) since this gives smaller error bars.
Towards this end the standard practice is to construct the interpolating fields
from `smeared' operators, see Appendix \ref{appRanSrc}.
In this way the time correlation functions
assume their asymptotic behaviour already at small $t$.

Another technique towards the same end was described in \cite{Lue90}, and adopted in
\cite{Goe94} for a lattice scattering problem. There, using the notation
of \cite{Goe94} for the moment, the generalized eigenvalue problem
${\cal C}(t){\bf w}^\nu = \lambda_\nu(t,t_0){\cal C}(t_0){\bf w}^\nu$ 
is considered. It leads to study the matrix 
${\cal D}(t,t_0)={\cal C}^{-\frac12}(t_0) {\cal C}(t) {\cal C}^{-\frac12}(t_0)$.
This is an interesting parallel to the form of (\ref{Ceff1}). It is stated in
\cite{Goe94} that a reliable determination of the energy levels already for
smaller values of $t$ is achieved.

\subsection{Lattice Symmetry}

From a numerical point of view computing ${\cal C}$ is greatly facilitated by 
utilizing symmetries of the lattice.
The action used with our $L^3\times T$
lattice is invariant under the group $O(3,\Bbb Z)$
of discrete transformations of the cubic sublattice.
Using common nomenclature \cite{Ham64} the irreducible representations are
$\Gamma=A_1^{\pm},A_2^{\pm},E^{+},T_1^{\pm},T_2^{\pm}$ with respective
dimensionalities $N_{\Gamma}=1,1,2,3,3$ .

Given a fixed discrete lattice momentum $\vec{p}\,'$, see (\ref{momenta}),
the application of all 
group transformations $g\in O(3,{\Bbb Z})$ generates a set of momenta
$\vec{p}={\cal O}_g\,\vec{p}\,'$ which all have the same length
$p=|\vec{p}\,'|$. These transformations define a representation of 
$O(3,{\Bbb Z})$, say its basis vectors are $|\vec{p}>$, which is in general
reducible. Let
\begin{equation}
|\vec{p}> = \sum_{\Gamma} \sum_{\epsilon} |(\Gamma,p)\epsilon> 
<(\Gamma,p)\epsilon|\vec{p}> \,,
\label{pRep}\end{equation}
with $|(\Gamma,p)\epsilon>$, $\epsilon=1\ldots N_{\Gamma}$, being a set
of basis vectors of the subspace
that belongs to $\Gamma$. Systematic construction of those basis vectors
is straightforward \cite{Ham64}. For example, choosing an on-axis momentum
$\vec{p}\,'=(p,0,0)$, and $\Gamma=A^+_1$, one obtains
\begin{eqnarray}
|(A^+_1,0)1> &=& |(0,0,0)> \quad \mbox{for}\;\; p=0 \\
|(A^+_1,p)1> &=& \frac{1}{\sqrt{6}} \sum_{\pm}
\left( |(\pm p,0,0)>+|(0,\pm p,0)>+|(0,0,\pm p)> \right) \nonumber \\
 & & \rule{19mm}{0mm} \quad \mbox{for}\;\; p>0 \,.
\label{A1p}\end{eqnarray}

It is also useful to know that representations $\underline{\ell}$ of the 
continuum group $O(3)$
characterized by their angular momentum $\ell=0,1\ldots 4$ have decompositions
\begin{eqnarray} 
\underline{0} &=& A^+_1 \nonumber \\ 
\underline{1} &=& T^-_1 \nonumber \\ 
\underline{2} &=& E^+ \oplus T^+_2 \nonumber \\ 
\underline{3} &=& A^-_2 \oplus T^-_1 \oplus T^-_2 \nonumber \\ 
\underline{4} &=& A^+_1 \oplus E^+ \oplus T^+_1 \oplus T^+_2 \label{A1l0} \,,
\end{eqnarray} 
if $\ell>4$ the last four lines apply cyclically.
The above follows from letting $O(3,{\Bbb Z})$ operate
on harmonic polynomials $r^\ell Y_{\ell m}(\Theta,\phi)$ which form a
basis for $\underline{\ell}$.

Since both the full and the free correlators $C^{(4)}$ and $\overline{C}^{(4)}$,
respectively, commute with all group operations $g\in O(d,{\Bbb Z})$
there exist reduced matrices, say $C^{(4;\Gamma)}$ and
$\overline{C}^{(4;\Gamma)}$, within each irreducible representation $\Gamma$
such that
\begin{equation}
<(\Gamma,p)\epsilon|C^{(4)}(t,t_0)|(\Gamma',q)\epsilon'> = 
\delta_{\Gamma\Gamma'}\delta_{\epsilon\epsilon'}
C^{(4;\Gamma)}_{p\,q}(t,t_0)
\label{C4Gam}\end{equation}
and similarly for $\overline{C}^{(4;\Gamma)}$.
It is obvious from (\ref{eq510}) and (\ref{eq511}) that the reduced matrix
elements of the free correlator have the form
\begin{equation}
\overline{C}^{(4;\Gamma)}_{p\,q}(t,t_0) = \delta_{p\,q}\,
\left| \bar{c}^{(\Gamma)}_{p}(t,t_0) \right| ^2 \, .
\label{eq73}\end{equation}
The functions $\bar{c}^{(\Gamma)}_{p}(t,t_0)$ are related to
$c_{\vec{p}}(t,t_0)$ by a $p$- and $\Gamma$-dependent factor.
The above representation of $\overline{C}^{(4;\Gamma)}$ is of great 
advantage for
numerical work because the inverse-square-root operation needed
to compute the effective correlator (\ref{Ceff1}) is now trivial.
In the sector $\Gamma$ we simply have
\begin{equation}
{\cal C}^{(4;\Gamma)}_{p\,q}(t,t_0) =
\frac{C^{(4;\Gamma)}_{p\,q}(t,t_0)}
{ \left| \bar{c}^{(\Gamma)}_{p}(t,t_0) \right|
  \left| \bar{c}^{(\Gamma)}_{q}(t,t_0) \right| }\,.
\label{eq75}\end{equation}
We envision numerical diagonalization, say
\begin{equation}
{\cal C}^{(4;\Gamma)}_{p\,q}(t,t_0) =
\sum_{n=1}^{N_\Gamma} \, v^{(\Gamma)}_n(p)
\lambda^{(\Gamma)}_n(t,t_0) \, v^{(\Gamma)}_n(q) \,.
\label{C4diag}\end{equation}
For asymptotic times the eigenvectors $v^{(\Gamma)}_n$ are time
independent \cite{Lue90}, whereas
the eigenvalues behave exponentially
\begin{equation}
\lambda^{(\Gamma)}_n(t,t_0) \sim
a^{(\Gamma)}_{n}\exp(-w^{(\Gamma)}_{n}(t-t_0)) \,.
\label{Wt}\end{equation}
It is worth noting that diagonalization on each time slice, if numerically
feasible, ensures that
excited states decouple from the ground state, and each other, by virtue of
orthogonal eigenvectors. Thus, at least for a large enough size of
${\cal C}^{(4;\Gamma)}$,
the $t\rightarrow \infty$ behavior avoids the two-body relative
ground state.

Extracting the \ERI\ as defined in (\ref{ERI}) now becomes a trivial matter.
Provided that the $v^{(\Gamma)}_n$ are orthonormal we have
\begin{equation}
{\cal H}_{I,pq}^{(\Gamma)} =
\sum_{n=1}^{N_\Gamma} \, v^{(\Gamma)}_n(p)
\, w^{(\Gamma)}_n \, v^{(\Gamma)}_n(q) \,.
\label{HIpq}\end{equation}
These are the desired matrix elements of the \ERI\ in the sector $\Gamma$.
In the basis $|\vec{p}>$ of lattice momenta we have
\begin{equation}
<\vec{p}\,|{\cal H}_I|\vec{q}> = \sum_{\Gamma} \sum_{\epsilon}
<\vec{p}\,|(\Gamma,p)\epsilon>
{\cal H}_{I,pq}^{(\Gamma)}
<(\Gamma,q)\epsilon|\vec{q}> \,.
\label{HIpRep}\end{equation}
How many irreducible representations
$\Gamma$ of the cubic lattice symmetry group actually are needed
in the above series depends on the hadron-hadron system under
consideration, the kinematical conditions, and on the
partial waves that couple into the total spin $J$ of the system.
In this context the decompositions (\ref{A1l0}) become relevant.

In the numerical part of this work only the sector $\Gamma=A_1^{+}$
is considered.

\subsection{Nonrelativistic Potentials}

From a kinematical point of view a $\pi$--$\pi$ system is
ill suited for a description by nonrelativistic potentials.
Despite to-be-expected strong relativistic effects
there are still good reasons, however, to pursue this venue. First, we
expect a potential to give insight into the nature of the residual
interaction (attractive, repulsive, range, etc.). Second, a lattice
simulation typically is performed
in the realm of large pseudoscalar mass, certainly larger than the
physical $\pi$ mass.
In this region the potential picture still may have some validity.
It is only through
extrapolation that the relativistic region will be reached. Third,
it seems that the shortage of lattice work in the area
of hadron-hadron interaction, particularly for real-world systems,
lends some justification to drastic approximations, at this stage.

The coordinate space matrix elements of the \ERI\ are obtained from
(\ref{HIpRep}) by discrete lattice Fourier transformation. 
After redefining momentum variables we have
\begin{equation}
<\vec{r}\,|{\cal H}_I|\vec{s}> =
L^{-3}\sum_{\vec{p}}\sum_{\vec{q}}
e^{ i\vec{p}\cdot(\vec{r}-\vec{s})}
e^{-i\vec{q}\cdot(\vec{r}+\vec{s})}
<\vec{p}-\vec{q}\,|{\cal H}_I|\vec{p}+\vec{q}> \,.
\label{HIrRep}\end{equation}
It is useful to write the matrix element of ${\cal H}_I$ as a sum of
$<-\vec{q}\,|{\cal H}_I|+\vec{q}>$ and a remainder. A reference system
where the relative momenta before and after a scattering event are $\pm\vec{q}$,
respectively, is known as the Breit frame \cite{Fes92}.
Then, the sum over $\vec{p}$ in (\ref{HIrRep}) gives rise to
$\delta_{\vec{r}\,\vec{s}}$ and thus to a local operator ${\cal V}(\vec{r}\,)$
for the Breit-frames contributions, and a genuinely nonlocal operator
${\cal W}(\vec{r},\vec{s}\,)$ for the remainder
\begin{equation}
<\vec{r}\,|{\cal H}_I|\vec{s}> =
\delta_{\vec{r}\vec{s}}\, {\cal V}(\vec{r}\,) + {\cal W}(\vec{r},\vec{s}\,) \,.
\label{HIVW}\end{equation}

Effectively, ${\cal V}$ and ${\cal W}$ comprise 
all microscopic dynamical effects, ultimately by way of the lattice action,
that possibly contribute to the residual interaction between the composite mesons.
Loosely speaking, these include potential as well as kinetic (hopping) effects.
In a quantum field theory this distinction hardly makes sense, however, there
exists an interesting parallel in the many-body theory of nuclear reactions.
There, techniques known as resonating group method (RGM) and the essentially equivalent
generator coordinate method (GCM) lead to effective interactions between composite
nuclei which are built from both kinetic and potential energies
of some microscopic Hamiltonian. The brief digression of Section \ref{secRGM}
will illuminate the analogy.

Following common semantics we will refer to ${\cal V}$ and ${\cal W}$ as potentials.
Their particular forms are
\begin{equation}
{\cal V}(\vec{r}\,) =
\sum_{\vec{q}} e^{-i2\vec{q}\cdot\vec{r}}
<-\vec{q}\,|{\cal H}_I|+\vec{q}>
\label{Vpot}\end{equation}
\begin{eqnarray}
{\cal W}(\vec{r},\vec{s}\,) &=&
L^{-3}\sum_{\vec{p}}\sum_{\vec{q}}
e^{ i\vec{p}\cdot(\vec{r}-\vec{s})}
e^{-i\vec{q}\cdot(\vec{r}+\vec{s})} \nonumber \\
 & & \rule{19mm}{0mm}
\left( <\vec{p}-\vec{q}\,|{\cal H}_I|\vec{p}+\vec{q}>
-<-\vec{q}\,|{\cal H}_I|+\vec{q}> \right) \,.
\label{Wpot}\end{eqnarray}
   
We will here limit ourselves to the relative S-wave local potential.
Projecting (\ref{Vpot}) onto the partial wave $\ell=0$,
\begin{equation}
{\cal V}_0(r) = \frac{1}{4\pi}\int d\Omega_{\vec{r}}\,{\cal V}(\vec{r}\,) \,,
\label{Vl0}\end{equation}
implies a {\em prescription for interpolation} between the lattice sites.
Whereas $\vec{r}$ in (\ref{Vpot}) is assumed discrete it has been
reinterpreted as a continuous variable for the purpose of integration in
(\ref{Vl0}). The effect of the angular integration on (\ref{Vpot}) is
that only diagonal $A^+_1$ sector reduced matrix elements of the expansion
(\ref{HIpRep}) survive. We have
\begin{equation}
{\cal V}_0(r) = \sum_q\,j_0(2qr)\,{\cal H}_{I,qq}^{(A^+_1)} \,,
\label{Vj0}\end{equation}
where $j_0(x)=\sin(x)/x$. Note that the sum in (\ref{Vj0}) runs over
discrete lattice momenta.

\subsubsection{Succinct RGM}\label{secRGM}

We divert to specify the analogies mentioned above.
The resonating group method (RGM), and it's cousin the generator coordinate method (GCM),
are specifically applicable to a system of two composite nuclei.
For the many-body wave function of two nuclei, say $1$ and $2$, the RGM ansatz is
$ \Phi = {\cal A}\left(\phi_1(\xi_1)\phi_2(\xi_2)g(\vec{x})\right)$,
where $\xi_{1,2}$ are intrinsic coordinates and $\vec{x}$ is the
{\em relative} coordinate between the clusters. The operator ${\cal A}$ antisymmetrizes
with respect to nucleon exchange. The microscopic Hamilton operator $H$,
with ${\cal A}H=H{\cal A}$, can be written as $H=H_1+H_2+T_{12}+V_{12}$ where $H_{1,2}$
refer to the intrinsic degrees of freedom and $T_{12}$, $V_{12}$ are the relative kinetic
and potential energy operators.
While $\phi_{1,2}$ are input, for example shell-model wave functions, the relative
wave function $g(\vec{x})$ is subject to solution of the RGM equation
$\langle\phi_1\phi_2\delta_{\vec{r}}|(H-E){\cal A}|\phi_1\phi_2g\rangle=0$,
all $\vec{r}$, where $\delta_{\vec{r}}(\vec{x})=\delta^3(\vec{x}-\vec{r})$.
In terms of the integral kernels
$n(\vec{r},\vec{r}\,') = \langle\phi_1\phi_2\delta_{\vec{r}}|{\cal A}|
\phi_1\phi_2\delta_{\vec{r}\,'}\rangle$
and
$h(\vec{r},\vec{r}\,') = \langle\phi_1\phi_2\delta_{\vec{r}}|H{\cal A}|
\phi_1\phi_2\delta_{\vec{r}\,'}\rangle$
the RGM equation reads
$(h-En)g=0$. Assuming that the so-called overlap kernel $n$ has no zero eigenvalues 
(redundant states) it can be cast into the form of a Schr{\"o}dinger equation. Defining
\begin{equation}
h_{RGM}=n^{-1/2}\,h\;n^{-1/2} \,,
\end{equation}
we have $(h_{RGM}-E)f=0$ with $f=n^{1/2}g$. Here $h_{RGM}=h_0+h_I$ is an effective
Hamiltonian for the two-nucleus system. Its free part, $h_0$, stems from splitting off
terms with no rest antisymmetrization, direct terms, in this case the first terms in
${\cal A} = {\Bbb 1}-({\Bbb 1}-{\cal A})$ and
$H{\cal A} = (H_1+H_2+T_{12})+V_{12}+H({\cal A}-{\Bbb 1})$
in the definitions of $n(\vec{r},\vec{r}\,')$ and $h(\vec{r},\vec{r}\,')$, respectively,
and also in $n^{-1/2}={\Bbb 1}+(n^{-1/2}-{\Bbb 1})$. Direct terms lead to local
operators on the wave function $f(\vec{r})$.
In this way $(h_{RGM}-E)f=0$ assumes the form
\begin{equation}
-\frac{\hbar^2}{2m_{12}}\Delta_{r}f(\vec{r}\,)
+V_{RGM}(\vec{r})\,f(\vec{r}\,) 
+\int d^3r\,' W_{RGM}(\vec{r},\vec{r}\,')\,f(\vec{r}\,') \\
= (E-E_1-E_2)f(\vec{r}\,) \,.
\label{RGM}\end{equation}
Here $m_{12}$ is the reduced mass, the kinetic operator is identical with $T_{12}$,
and $E_{1,2}$ are the internal cluster energies.
Microscopic dynamics with kinetic and potential energies, contained in $H$,
are cast into the effective potentials $V_{RGM},W_{RGM}$.

Analogies with (\ref{Heff}) and (\ref{HIVW}) suggest themselves.
Since the spectrum of ${\cal H}_0$ is akin to $2\sqrt{m^2+p^2}=2m+p^2/m+\ldots$,
similarities exist between $E_1+E_2-\frac{\hbar^2}{2m_{12}}\Delta_{x}$
and ${\cal H}_0$, and between $V_{RGM},W_{RGM}$ and ${\cal V},{\cal W}$ or ${\cal H}_I$.
Loosely speaking $2m$ plus the relative kinetic energy of our two-pion system drops
out in the ratio of (\ref{Ceff1}).

\section{Computational Issues, Analysis and Results}\label{secComp}

The numerical implementation poses some
extraordinary challenges. Compared to typical hadronic mass
scales, energy shifts due to residual interaction can be quite small.
On the other hand, the total energies of a two-hadron system
at larger relative momenta tend to be big.
Thus extracting energy shifts from ratios, see
(\ref{eq75}), of steeply dropping time correlation functions can push
the numerical quality of the lattice simulation to its limits.

\subsection{Lattice Implementation}\label{secLat}

The computation has been done on an $L^3\times T = 9^3\times 13$ lattice.
We have used a tree-level $O(a^2)$ tadpole improved next-nearest-neighbor
action, as discussed in \cite{Egu84,Ham83a}, in a quenched simulation.
The number of gauge configurations was $N_U=208$ separated by at least 1024
update steps.  The hadron spectrum of this action has been studied in
\cite{Fie96d}, and also in \cite{Lee98}, on coarse lattices.
Quite good agreement with the experimental masses was observed in both cases.
In the conventions of \cite{Fie96d} the gauge field
coupling parameter was $\beta=6.2$. This corresponds to a lattice
constant of
\begin{equation}
a \simeq 0.4\,\mbox{fm} \quad \mbox{or} \quad a^{-1} \simeq 0.5\,\mbox{GeV}
\label{afmMeV}\end{equation}
as determined from the string tension\footnote{
Unless otherwise noted all lengths and energies are given in units
of $a$ and $a^{-1}$, respectively, throughout the text and the figures.}.
This lattice is rather coarse,
on the other hand a large lattice volume (here $La\simeq 3.6\mbox{fm}$)
is required to accommodate two hadrons and to allow for sufficiently
large separations at which their residual interaction becomes negligible.

The fermion part of the improved action \cite{Fie96d} contains
the Wilson hopping parameter $\kappa$. Quark propagators were obtained
for the six values
\begin{equation}
\kappa^{-1}= 5.720,\, 5.804,\, 5.888,\, 5.972,\, 6.056,\, 6.140 \,,
\label{kappas}\end{equation}
using a multiple mass algorithm \cite{Gla96}.
The critical value for $\kappa^{-1}$ is about $5.5$.
The correlator matrix (\ref{eq46}) requires quark propagator
matrix elements $G(\vec{x}t,\vec{y}t_0)$
between arbitrary lattice
sites, except that the initial time slice is fixed. 
Computing all of those is not feasible.
A random source technique very similar to the one described
in \cite{Fie94a} was employed, see Appendix \ref{appRanSrc}.
The number of complex Gaussian random sources was $N_R=8$ for each
color-Dirac source point, thus we effectively have
$8\times 3\times 4=96$ random sources per gauge configuration available to
estimate $G$.

The fermion fields were subjected to `Gaussian smearing'. 
We have used the smearing strength parameter $\alpha=2$, in the notation 
of \cite{Ale94}, and $S=4$ iterations for smearing at the sink. 
This parameter set is similar to the one used in \cite{Fie96d}
with the same type action.
In addition `APE fuzzing' \cite{Alb87a}, which is the analogue of
smearing for link variables, was done on each gauge configuration
prior to smearing, also with $\alpha=2$ and $S=4$.

\subsection{Correlator Analysis}

The centerpiece of the numerical effort is the effective correlator
matrix (\ref{eq75}). We have computed $5\times 5$ $A^+_1$-sector matrices
$C^{(4;A^+_1)}(t,t_0)$ and $\overline{C}^{(4;A^+_1)}(t,t_0)$
for on-axis momenta 
\begin{equation}
q=\frac{2\pi}{L}k \quad\mbox{with}\quad k=0,1,2,3,4 \,.
\label{kmax}\end{equation}
The source time slice is at $t_0=3$. It turns out that the
$C^{(4;A^+_1)}(t,t_0)$ are dominated by their diagonal elements.
The off-diagonal elements are small and have somewhat large statistical
errors, mostly extending across zero.
\begin{figure}[hbt]
\begin{center}
\mbox{\epsfxsize=78mm\rotate[l]{\epsfbox{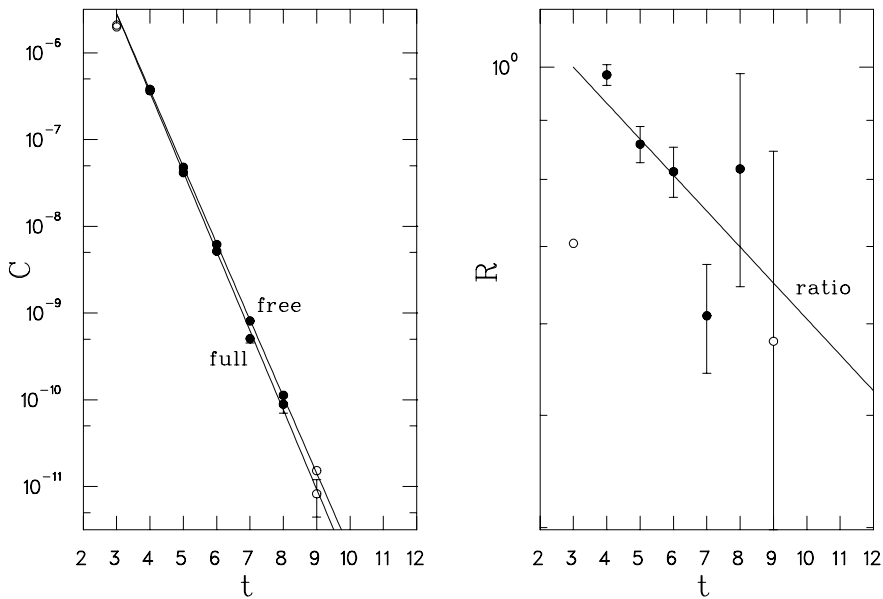}}}
\end{center}
\caption{Free and full correlator functions (C) and their ratio (R),
for $\kappa^{-1}=5.720$ and momentum $k=1$, showing a repulsive mode.}
\label{figP1K1}\end{figure}
\begin{figure}[htb]
\begin{center}
\mbox{\epsfxsize=78mm\rotate[l]{\epsfbox{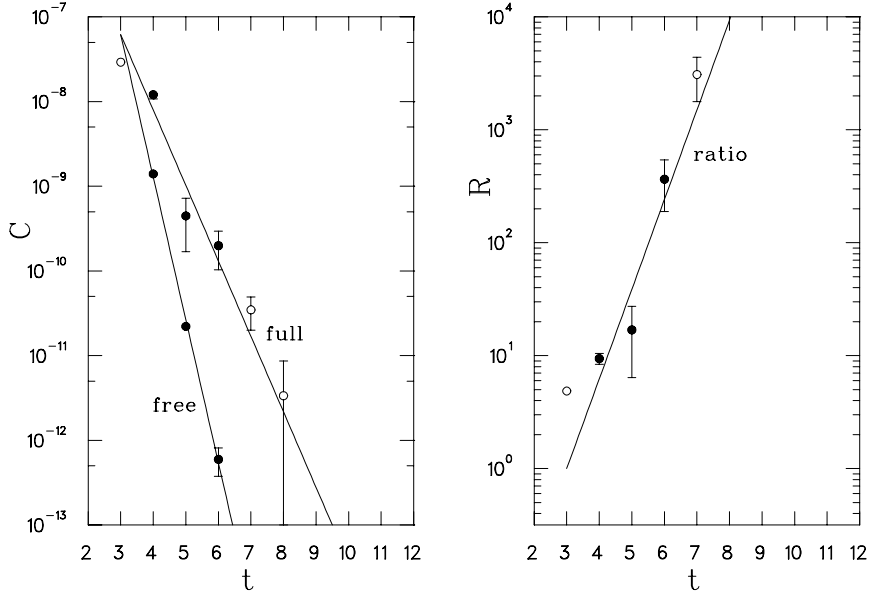}}}
\end{center}
\caption{Free and full correlator functions (C) and their ratio (R),
for $\kappa^{-1}=5.720$ and momentum $k=3$, showing an attractive mode.}
\label{figP3K1}\end{figure}
Hence it is reasonable to
adopt the diagonal approximation, replacing
\begin{equation}
C^{(4;A^+_1)}_{p\,q}(t,t_0) \simeq
\delta_{p\,q}\,C^{(4;A^+_1)}_{q\,q}(t,t_0) \,.
\label{diagC4}\end{equation}
Thus the effective correlator becomes
\begin{equation}
{\cal C}^{(4;A^+_1)}_{q\,q}(t,t_0) \simeq
\frac{C^{(4;A^+_1)}_{q\,q}(t,t_0)}
{\overline{C}^{(4;A^+_1)}_{q\,q}(t,t_0) } \,,
\label{diagCC}\end{equation}
with off-diagonal elements set to zero.
 
In Figs.~\ref{figP1K1} and \ref{figP3K1}
we show two examples of free and full correlator functions and their
ratios, as they appear in (\ref{diagCC}), both for the smallest
inverse hopping parameter, $\kappa^{-1}=5.720$.
The examples are for momenta $k=1$ and $k=3$,
respectively, see (\ref{kmax}). For the repulsive level the mass
shift is quite small, which translates into a very noisy signal 
for the ratio. For the attractive level the mass shift is rather
large, however, the fall-off of the free correlator function is
very steep, which has the effect that only very few time slices
are available for analysis. The straight lines in
Figs.~\ref{figP1K1} and \ref{figP3K1} come from linear fits to
the logarithms of the correlator functions. Only data points marked
by filled plot symbols were used in the fits. 

The lattice action used here, see Sect.~\ref{secLat}, exhibits ghosts.
These are unphysical branches in the lattice-quark dispersion relation
and are indigenous to highly improved actions \cite{Alf97}.
The presence of ghosts can contaminate the signal on early time
slices. In our case this contamination is clearly discernible only
at $t=3$. The corresponding data points of the correlator functions
have been ignored in our analysis.

Quantum fluctuations manifest themselves in an overall renormalization factor
that relates the free and full correlators at $t=t_0$
\begin{equation}
C^{(4;A^+_1)}_{q\,q}(t_0,t_0) = Z^4 \,
\overline{C}^{(4;A^+_1)}_{q\,q}(t_0,t_0) \,.
\label{Z4}\end{equation}
The value obtained from the fits is $Z^4=1.97(3)$, or $Z\simeq 1.18$. It 
is the same for all $\kappa$, a consequence of the quenched approximation.
We have plotted $Z^{-4} C^{(4;A^+_1)}_{q\,q}(t,t_0)$ in 
Figs.~\ref{figP1K1} and \ref{figP3K1} to
improve readability. Note that $Z^4$ has no influence on ${\cal H}_I$. 

\subsection{Residual Interaction}

\begin{figure}[htb]
\begin{center}
\mbox{\epsfxsize=60mm\epsfbox{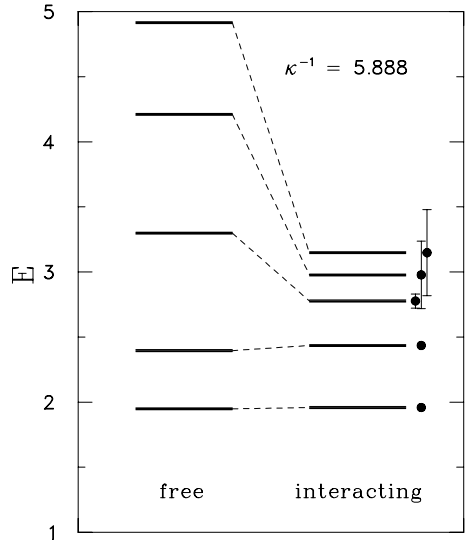}}
\end{center}
\caption{Energy level shifts (hyperfine splittings) due to the
residual interaction for the momenta $k=0,1,2,3,4$ at a fixed $\kappa^{-1}$.}
\label{figEK3}\end{figure}

In Fig.~\ref{figEK3}
we show an example of the energy level shifts of the
meson-meson system due to the residual interaction, for a selected $\kappa$.
Low-momentum
(long-distance) modes are weakly repulsive, whereas high-momentum
(short-distance) modes are strongly attractive. Errors stem from a bootstrap
analysis.

\begin{figure}[htb]
\begin{center}
\mbox{\epsfxsize=70mm\epsfbox{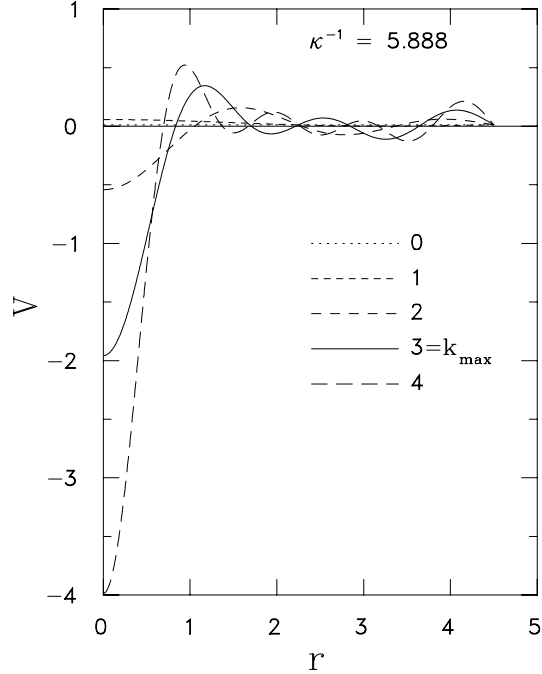}}
\end{center}
\caption{Fourier-Bessel representation (\protect\ref{Vj0}) of the S-wave
local potential, at $\kappa^{-1}=5.888$. Various momentum truncations,
see (\protect\ref{qtok}), are shown.}
\label{figVK3}\end{figure}

Analyzing the effective correlator (\ref{diagCC}) along the lines
(\ref{C4diag})-(\ref{HIpq})
yields the matrix elements ${\cal H}_{I,qq}^{(A^+_1)}$. 
These were used to compute S-wave
local potentials ${\cal V}_0(r)$ according to (\ref{Vj0}).
For the current lattice the sum over on-axis momenta (\ref{kmax}),
\begin{equation}
\sum_q \rightarrow \sum_{k=0}^{k_{\max}} \,,
\label{qtok}\end{equation}
truncates at $k_{\max}=4$.
We try to obtain some feel for
systematic errors caused by the momentum cut-off $k_{\max}$. On a coarse lattice
one should expect those to be large. Thus, in Fig.~\ref{figVK3}
we show a family of potentials, all for
$\kappa^{-1}=5.888$, which correspond to various upper limits $k_{\max}$.

\begin{figure}[htb]
\begin{center}
\mbox{\epsfxsize=59mm\rotate[l]{\epsfbox{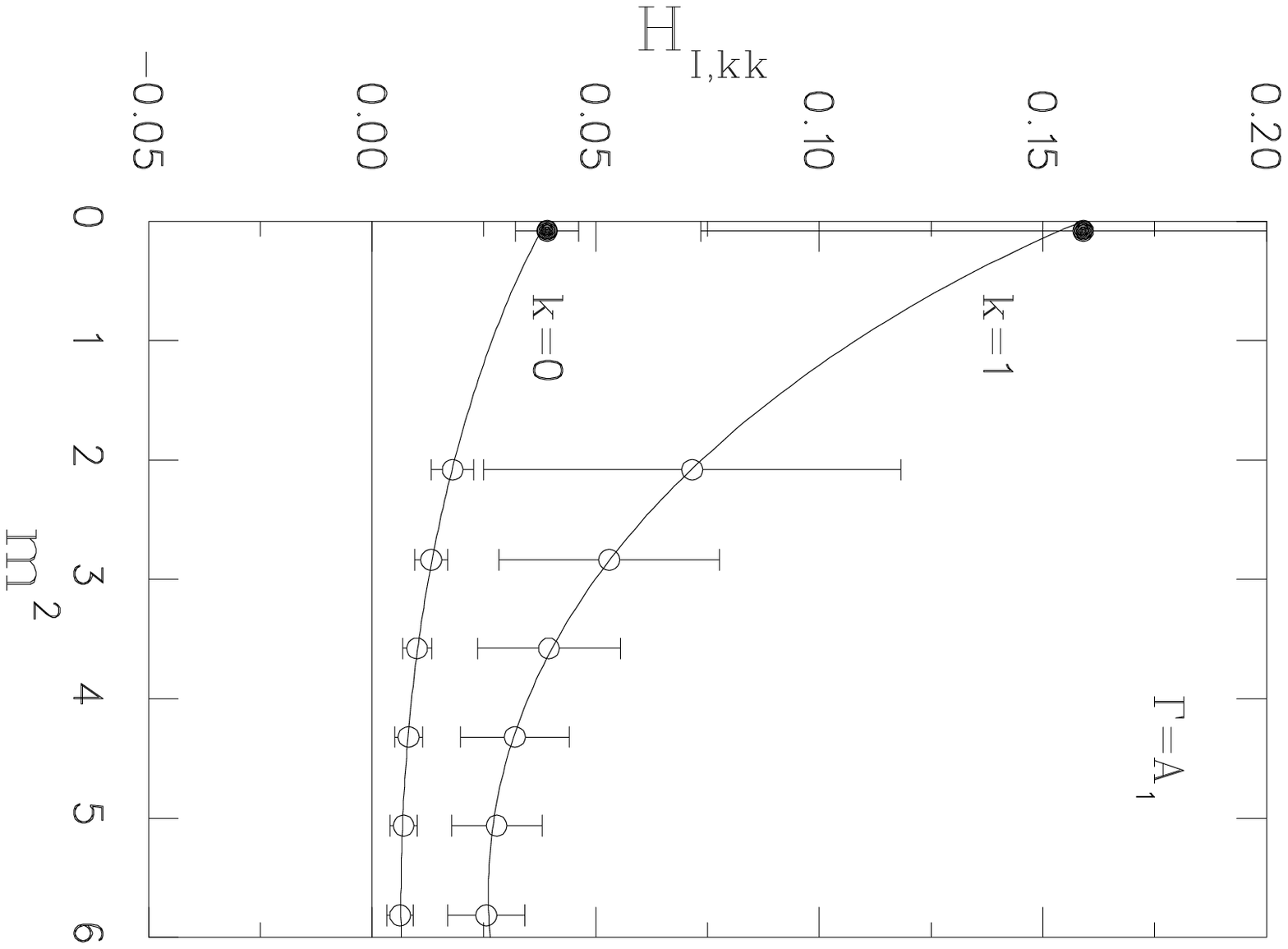}}\rule{2mm}{0mm}
      \epsfxsize=59mm\rotate[l]{\epsfbox{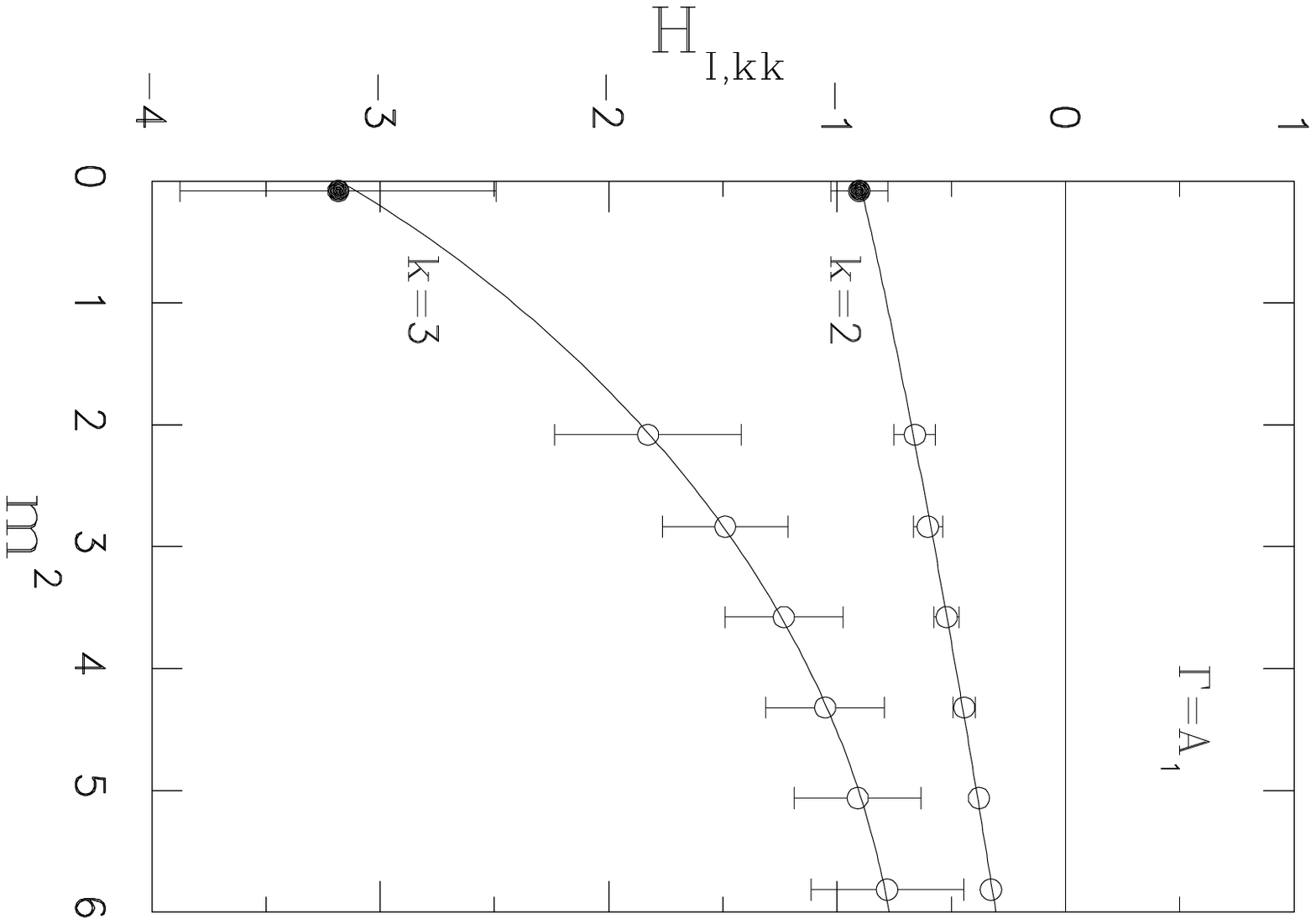}}\rule{2mm}{0mm}
      \epsfxsize=59mm\rotate[l]{\epsfbox{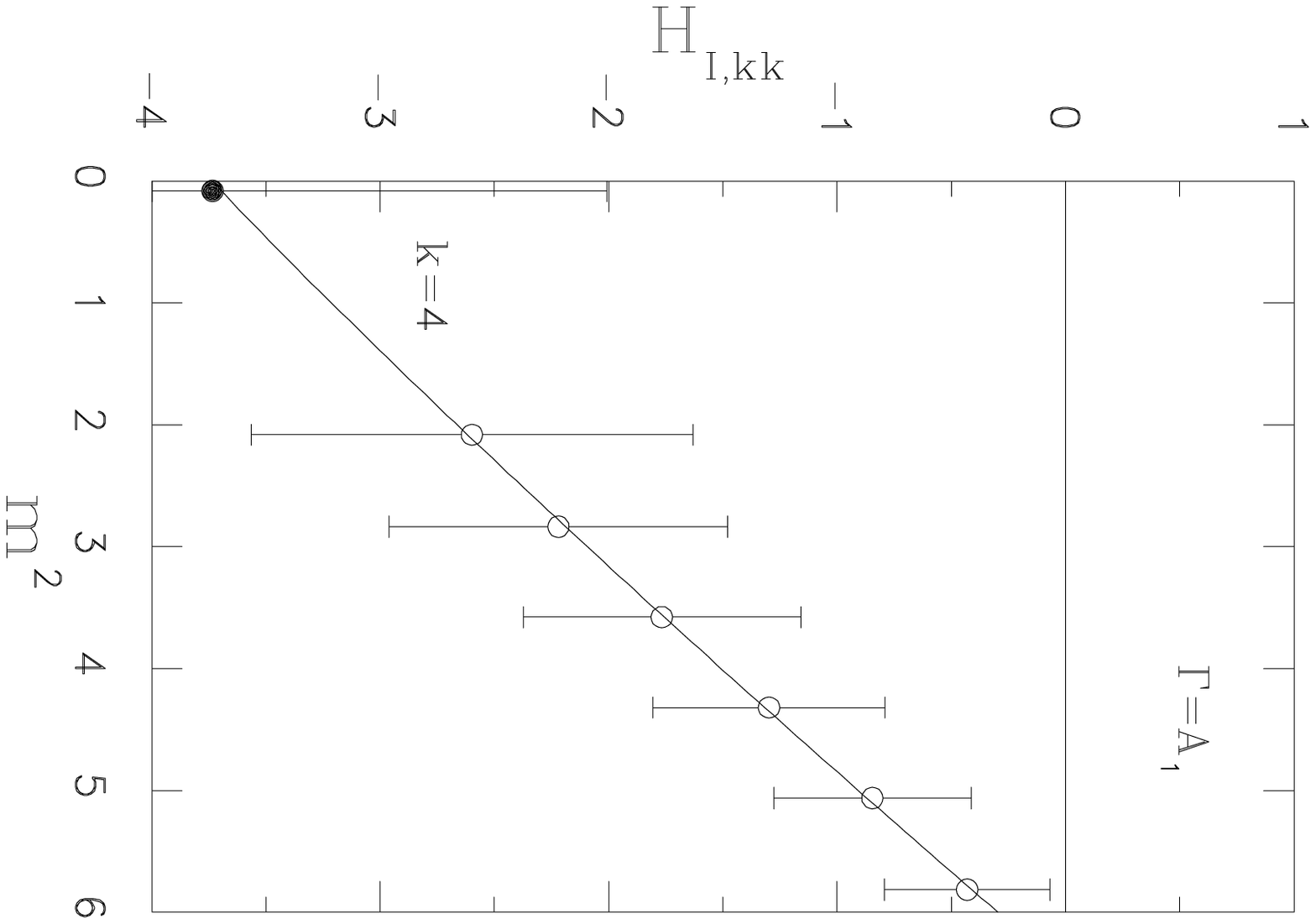}}}
\end{center}
\caption{Chiral extrapolations ${\cal H}_{I,qq}^{(A^+_1)}(m^2\rightarrow 0)$
for momenta $k=0\ldots 4$, see (\protect\ref{kmax}). In physical units
the six masses $m$ range from 1.2 to 0.72 GeV, the filled plot symbols
refer to $m_\pi$=0.14 GeV.}
\label{figCXH01234}\end{figure}

The matrix elements of the residual interaction (\ref{HIpq}) depend on
the pseudoscalar mass $m$ through the hopping parameter $\kappa$.
Figure~\ref{figCXH01234}
shows ${\cal H}_{I,qq}^{(A^+_1)}$ as a function
of $m^2$ for momenta $k=0\ldots 4$. The solid lines are three-parameter
fits using
\begin{equation}
h_q(x)=h_q+h_q'x+h_q''x^{3/2} \quad \mbox{with} \quad x=m^2 \,,
\label{chirH}\end{equation}
which is motivated by chiral perturbation theory \cite{Lab94}.
Extrapolation to the chiral limit
\begin{equation}
{\cal H}_{I,qq}^{(A^+_1)} = h_q(x\rightarrow 0)
\label{chirX}\end{equation}
then yields matrix elements of ${\cal H}_{I}$ which describe the residual
interaction of the physical $\pi$--$\pi$ system in the $I=2$ channel.
We will exclusively refer to those in the subsequent discussion.

\subsection{Parameterization}

\begin{figure}[htb]
\begin{center}
\mbox{\epsfxsize=60mm\epsfbox{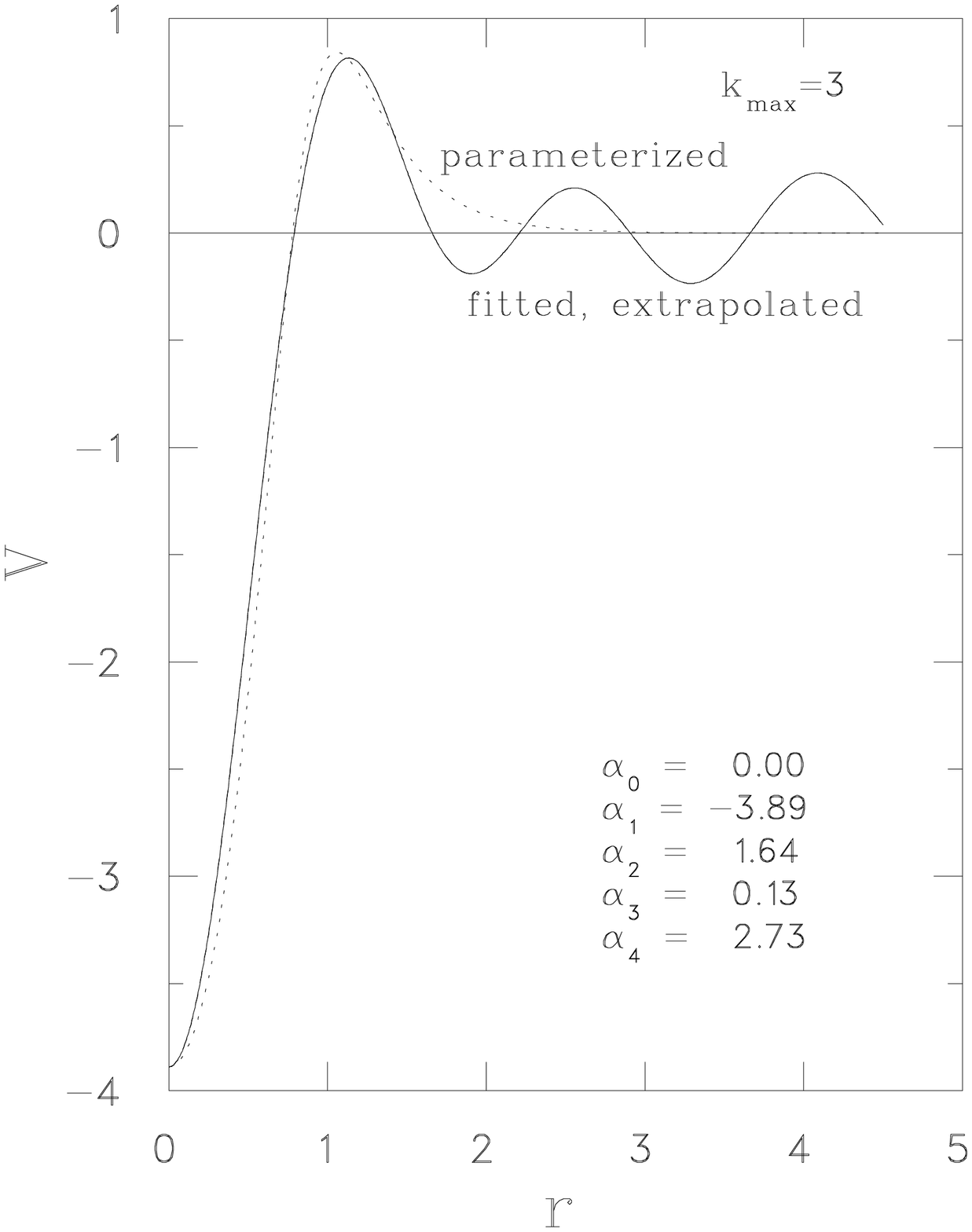}\rule{5mm}{0mm}
      \epsfxsize=63mm\epsfbox{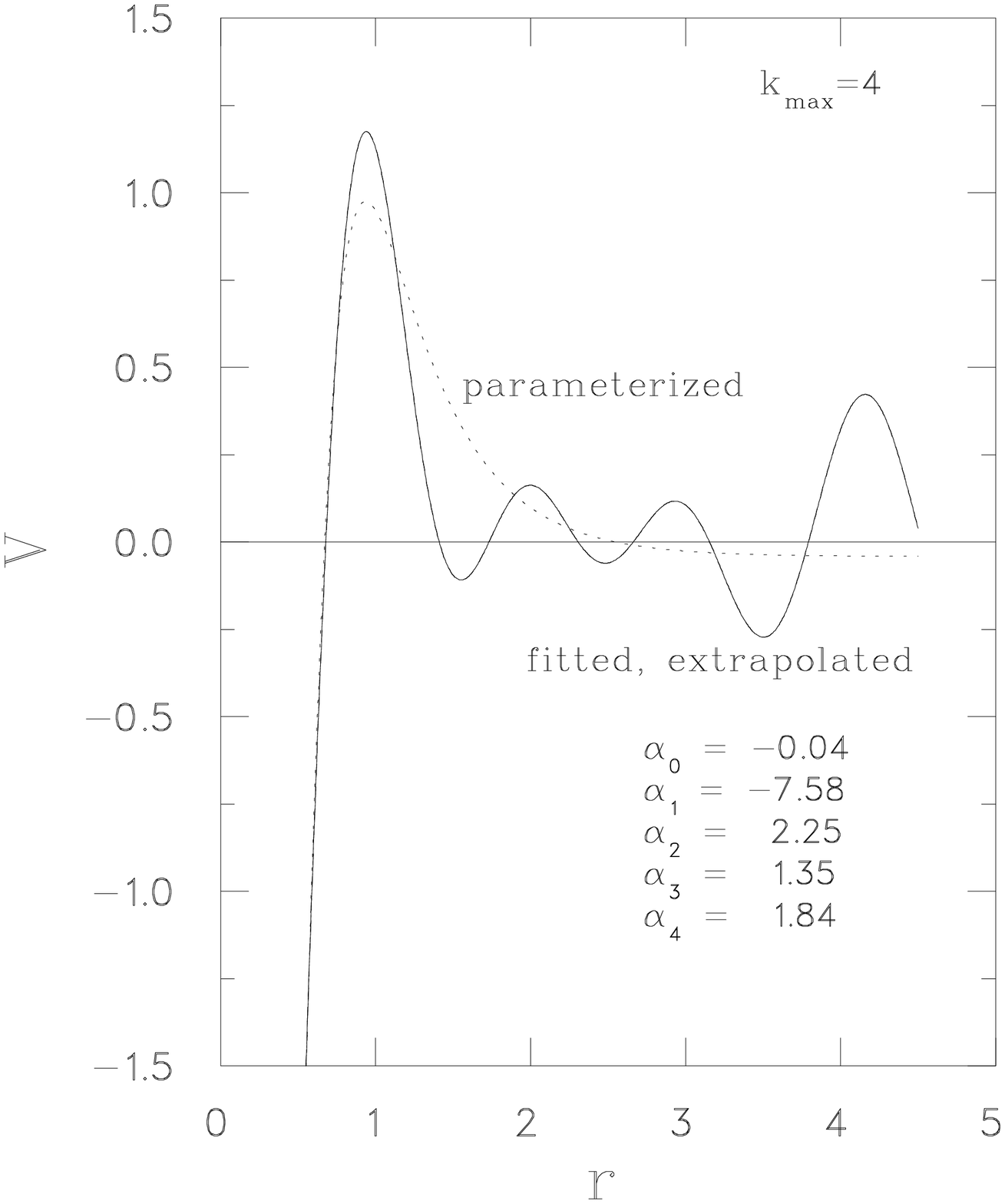}}
\end{center}
\caption{Local potentials for momentum truncations $k_{\max}=3$ and 
$k_{\max}=4$. Shown are the raw lattice results in the $m^2\rightarrow 0$
limit (extrapolated, solid line), the fits with the Fourier-Bessel trial
potential $V_L^{(\alpha)}(r)$ (fitted, solid line),
and the corresponding $r\in[0,\infty)$ parameterization
$V^{(\alpha)}(r)$ (parameterized, dashed line).}
\label{figVl}\end{figure}

In Fig.~\ref{figVl}
the solid lines correspond to the 
extrapolated potentials for $k_{\max}=3$ and $4$, respectively. The
oscillations caused by the representation as a truncated Fourier-Bessel
series complicate their interpretation.

To get a clearer picture, consider the parametric function
\begin{equation}
V^{(\alpha)}(r) = \alpha_1
\frac{ 1-\alpha_2 r^{\alpha_5} }
{ 1+\alpha_3 r^{\alpha_5+1} \exp(\alpha_4 r)} + \alpha_0 
\quad \mbox{where}\quad r\in [0,\infty) \,.
\label{Valpha}\end{equation}
It is flexible enough to represent attraction and/or repulsion at 
certain ranges, and has the feature that the Yukawa form
is approached asymptotically,
\begin{equation}
V^{(\alpha)}(r) \rightarrow
-\frac{\alpha_1\alpha_2}{\alpha_3}
\frac{e^{-\alpha_4 r}}{r} + \alpha_0 \quad \mbox{for}\quad r\rightarrow \infty \,.
\label{Vlimits}\end{equation}
Now define a `latticized' version of (\ref{Valpha}) via
\begin{equation}
V_L^{(\alpha)}(r) = \sum_{k=0}^{k_{\max}}\, j_0(2q_k r)\, V_k^{(\alpha)}
\quad\mbox{with}\quad q_k=\frac{2\pi}{L}k \,.
\label{VLalpha}\end{equation}
The $V_k^{(\alpha)}$ are expansion coefficients which we have determined
from calculating the matrix $B_{lk}=j_0(2q_k r_l)$, with support on 
$r_0=0$, $r_l=\frac{L}{4(k_{\max}-l+1)}$, $l=1\ldots k_{\max}$,
and then applying its inverse to (\ref{Valpha}) as
\begin{equation}
V_k^{(\alpha)} = \sum_{l=0}^{k_{\max}}\, B^{-1}_{kl} V^{(\alpha)}(r_l) \,.
\label{VKalpha}\end{equation}
In this way the span of $ V_L^{(\alpha)}(r)$ is a subspace of the space
of functions ${\cal V}_0(r)$, see (\ref{Vj0}) and (\ref{qtok}),
accessible through the lattice simulation.
Now make fits to lattice potentials minimizing
\begin{equation}
\int_0^{L/2}dr\,\left|{\cal V}_0(r)-V_L^{(\alpha)}(r)\right|^2 =
\min(\alpha)\,.
\label{VLmin}\end{equation}
We have found that $\alpha_5=2$ gives an almost perfect match and thus
held this parameter fixed. 
Also, $\alpha_0$ deviates very little from $0$ which indicates that the
interaction `around the world', across $La=3.6\mbox{fm}$, is negligible.
The results of the fits are the curves in Fig.~\ref{figVl} marked
`fitted'. In fact the `fitted' and the `extrapolated'
curves are indistinguishable within the line thickness.
(This changes if $\alpha_5\ne 2$.)
The curves marked `parameterized' show the corresponding $r\in[0,\infty)$
parameterization $V^{(\alpha)}(r)$ as defined in (\ref{Valpha}).

\subsection{Physical Potential}

\begin{figure}[htb]
\begin{center}
\mbox{\epsfxsize=37mm\epsfbox{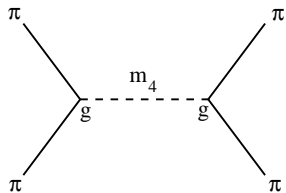}}
\end{center}
\caption{Boson exchange diagram related to (\protect\ref{Vlimits}) and
(\protect\ref{BosXm}),(\protect\ref{BosXg}).}
\label{figBosX}\end{figure}

\begin{figure}[htb]
\begin{center}
\mbox{\epsfxsize=60mm\epsfbox{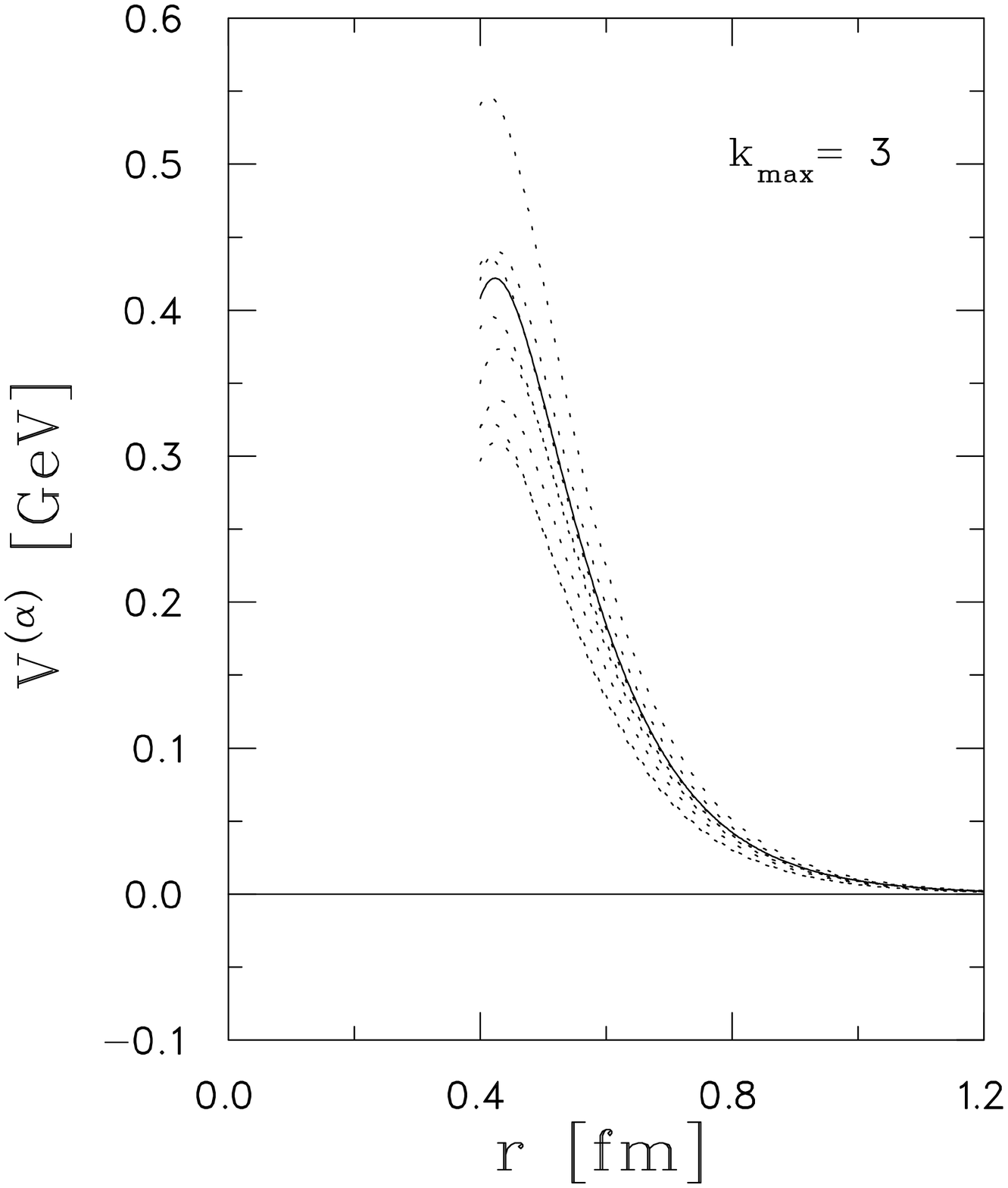}\rule{5mm}{0mm}
      \epsfxsize=60mm\epsfbox{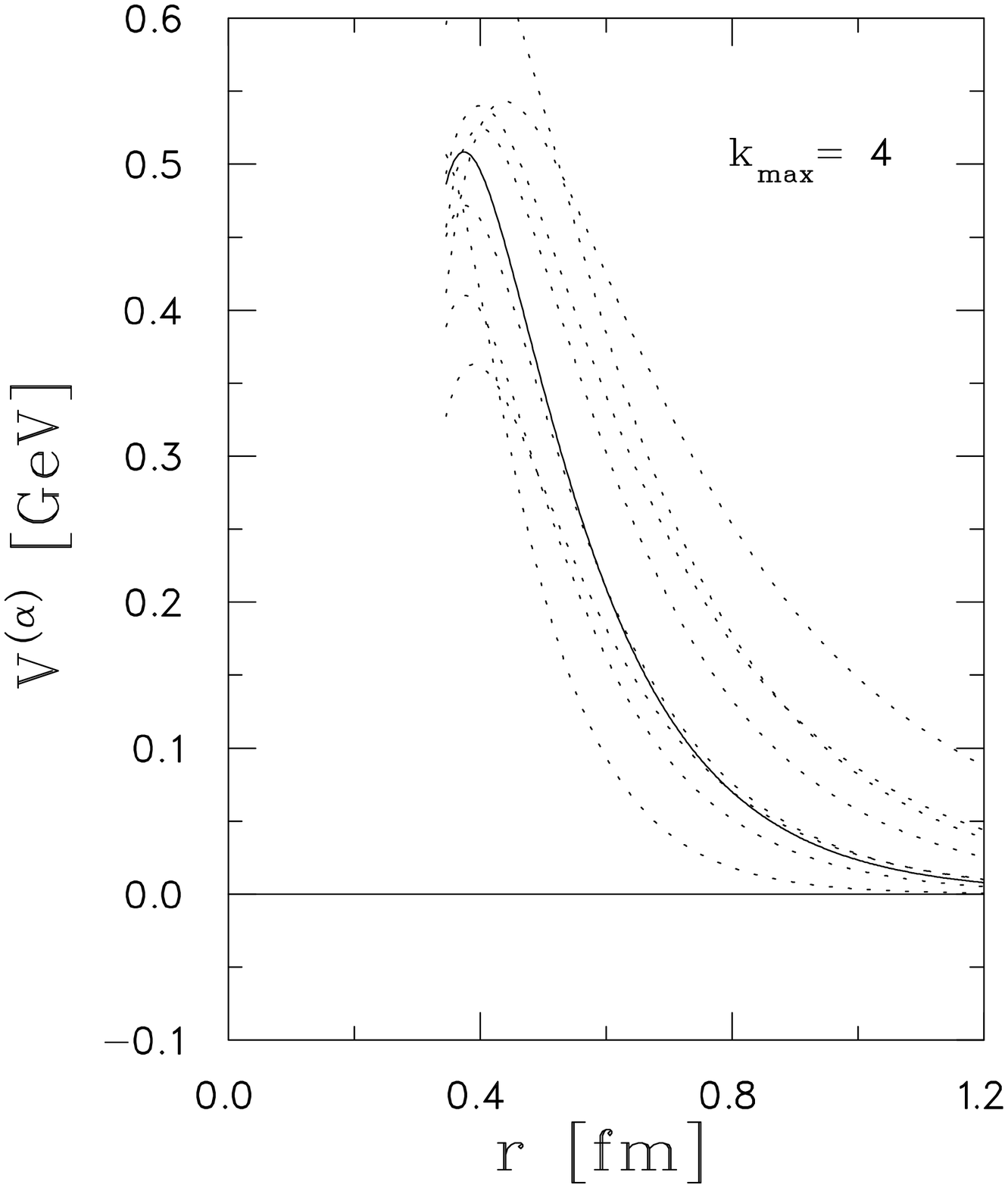}}
\end{center}
\caption{Local potentials for momentum truncations $k_{\max}=3$ and 
$k_{\max}=4$ (solid lines) and families of bootstrap samples (dashed lines)
to estimate the overall statistical errors.}
\label{figVe}\end{figure}

It is tempting to use the resulting fit parameters, shown in 
Fig.~\ref{figVl}, to make contact with the boson-exchange picture
of strong interactions, see Fig.~\ref{figBosX}.
This can be done identifying the Yukawa
asymptotics (\ref{Vlimits}) of $V^{\alpha}(r)$ with
\begin{eqnarray}
m_4&=&\alpha_4\,a^{-1} \simeq \, 1.4 \mbox{--} 0.9\,\mbox{GeV}  \label{BosXm} \\
\frac{g^2}{4\pi} &=& -\frac{\alpha_1\alpha_2}{\alpha_3}
\simeq \, 17.4 \mbox{--} 2.5 \,,\label{BosXg}
\end{eqnarray}
see (\ref{afmMeV}). The left and right numbers above relate 
to $k_{\max}=3$ and $k_{\max}=4$, respectively,
indicating large systematic errors.
Both the mass $m_4$ of the exchanged particle
and the vertex coupling, $g\simeq 14.8 \mbox{--} 5.6$, are typical
for a hadronic system. For example $g$ is around 13.45 for the $\pi$--N
vertex that enters the N--N interaction.
Statistical errors are also large for these
quantities (see Fig.~\ref{figVe}).

\begin{figure}[htb]
\begin{center}
\mbox{\epsfxsize=80mm\epsfbox{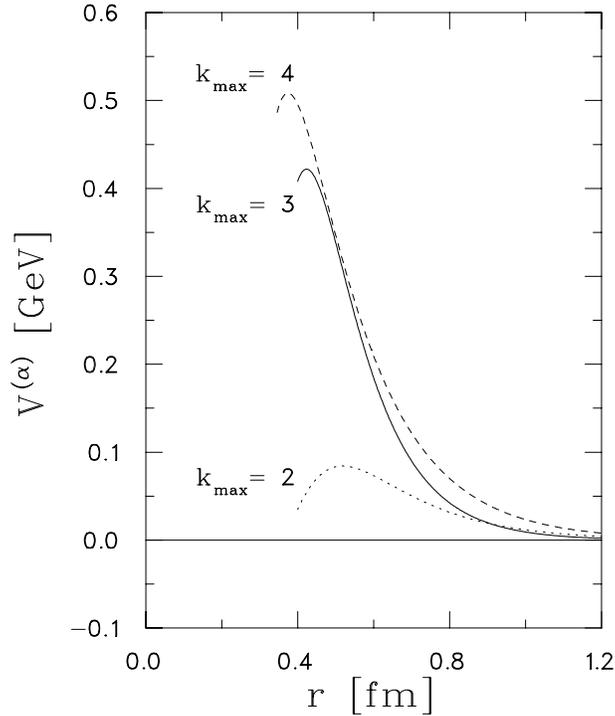}}
\end{center}
\caption{Local potentials for $k_{\max}=2,3,4$ in physical units for
distances $r<0.4\mbox{fm}$.}
\label{figV}\end{figure}

The above discussion addresses the long-range physics of the system.
Figure~\ref{figV}
shows the potentials in the region beyond the lattice
cut-off, $r\gtrsim a$, in physical units. 
Both truncations at $k_{\max}=3,4$ are in agreement and
indicate a repulsive residual interaction. In Fig.~\ref{figVe}
we show the physical potentials for truncations
$k_{\max}=3,4$, respectively, together with a family of potentials
computed from 8 bootstrap samples. The spread of the dashed lines in
Fig.~\ref{figVe} is indicative of the statistical fluctuations.
Compared to the $k_{\max}=3$ case
the fluctuations of the $k_{\max}=4$ potential are visibly larger.
A generic feature of two-hadron systems comes to mind as a possible reason.
Correlator matrix elements which involve large
momenta describe very massive (two-body) states and
consequently drop very steeply with $t$.
On small isotropic lattices it is very hard to deal with the
resulting deterioration of the lattice signal since data from only
very few time slices are usable.
From a numerical point of view we consider our
results using the truncation $k_{\max}=3$ the most reliable.
Nevertheless, we observe that both momentum cut-off
values lead to qualitatively consistent potentials in the
long and intermediate range region, $r\gtrsim 0.4\mbox{fm}$.

The steep drop of the potentials from their values at $r=a$ to 
deeply negative values at $r=0$, see Fig.~\ref{figVl}, 
does not influence low-energy physics. It is an indication, however, that
the physics in the system at $r=0$ is special.
A possible explanation is suggested by the $SU(3)$ color content of the
two-body system. Using standard nomenclature \cite{Lic78,Clo79} we
note that for the color structure of the one-meson interpolating field
(\ref{onephi}) only the singlet from
$\overline{\bf 3} \otimes \bf 3 = \bf 8 \oplus \bf 1$ is used. 
Thus the color-source structure of the two-meson
interpolating field (\ref{twophi}) is that of an overall singlet,
$\bf 1 \otimes \bf 1  = \bf 1$. A singlet is also contained in the
decomposition of the product of two color octets
$\bf 8 \otimes \bf 8 = \bf 27 \oplus \bf 10 \oplus \bf 8 \oplus \bf 8
\oplus \overline{\bf 10} \oplus \bf 1$.
Naive gluon exchange in $\bf 8 \otimes \bf 8$ is attractive \cite{Clo79}.
However, in the confinement phase the propagation of the
system into an $\bf 8 \otimes \bf 8$ color configuration
is dynamically suppressed with increasing distance $r$.
This means that we should expect the interaction
energy to be more attractive at $r=0$ as opposed to all
other $r=a,2a,3a\ldots$.
The situation is of course more complicated because of dynamical effects
from the spin (Dirac) and the flavor degrees of freedom.

In simulations of heavy-light meson-meson systems performed in coordinate space
\cite{Mih97} the above mechanism is seen directly. There, the $r=0$ case 
can be easily isolated. In our Fourier-Bessel analysis, on the other hand,
strong attraction at $r= 0$ can bias the result at $r\simeq a$ 
via oscillations of
the basis functions. The danger of course is that oscillatory features
are misinterpreted as repulsion. We have tried to minimize this by
employing the parameterization (\ref{Valpha}) which would easily be capable
of revealing attraction, say at $r=a$, but in fact gives a repulsive
interaction in the intermediate and long distance region, $r\gtrsim a$, as an
answer.

\subsection{Scattering Phase Shifts}
 
\begin{figure}[htb]
\begin{center}
\mbox{\epsfxsize=72mm\epsfbox{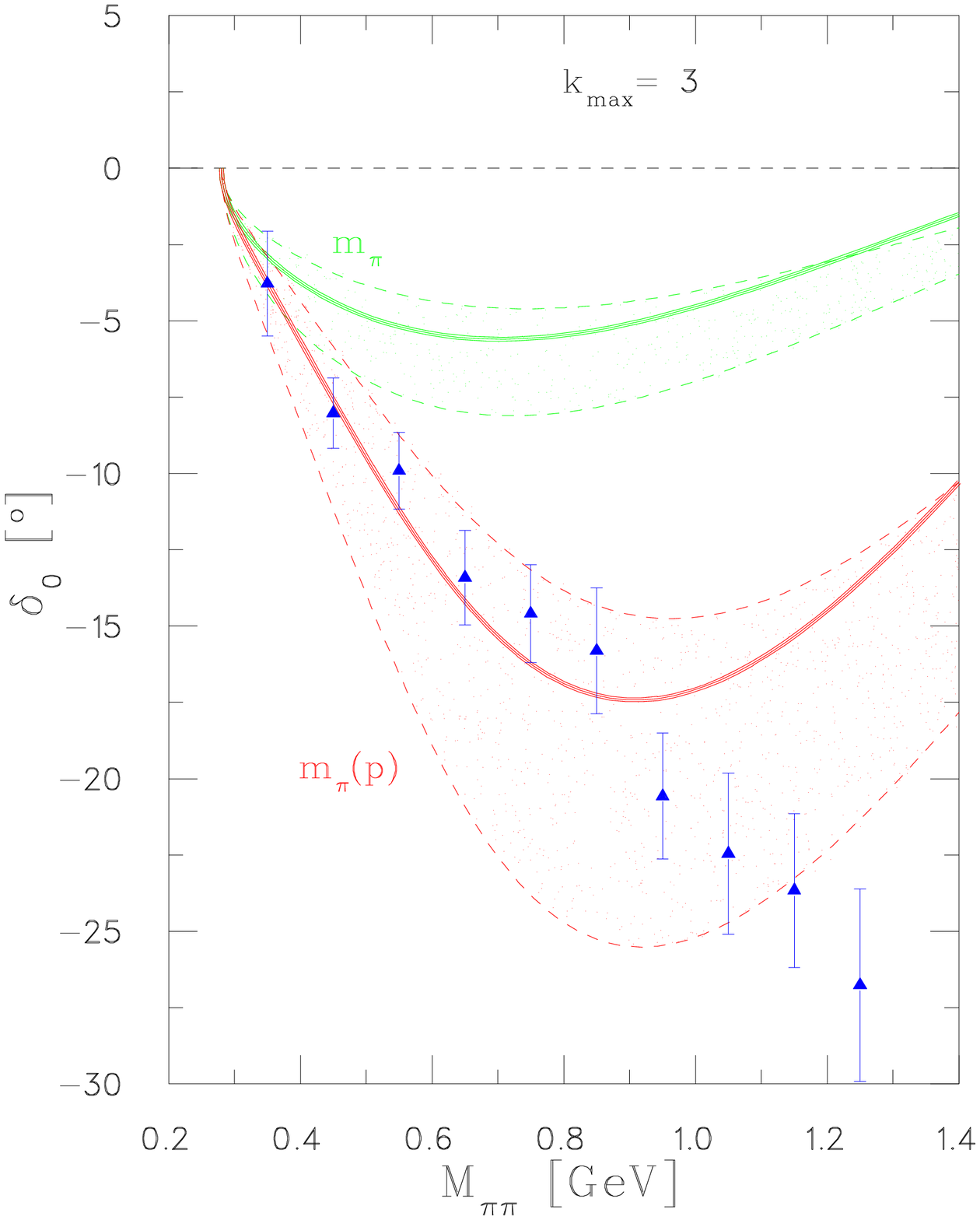}\rule{5mm}{0mm}
      \epsfxsize=72mm\epsfbox{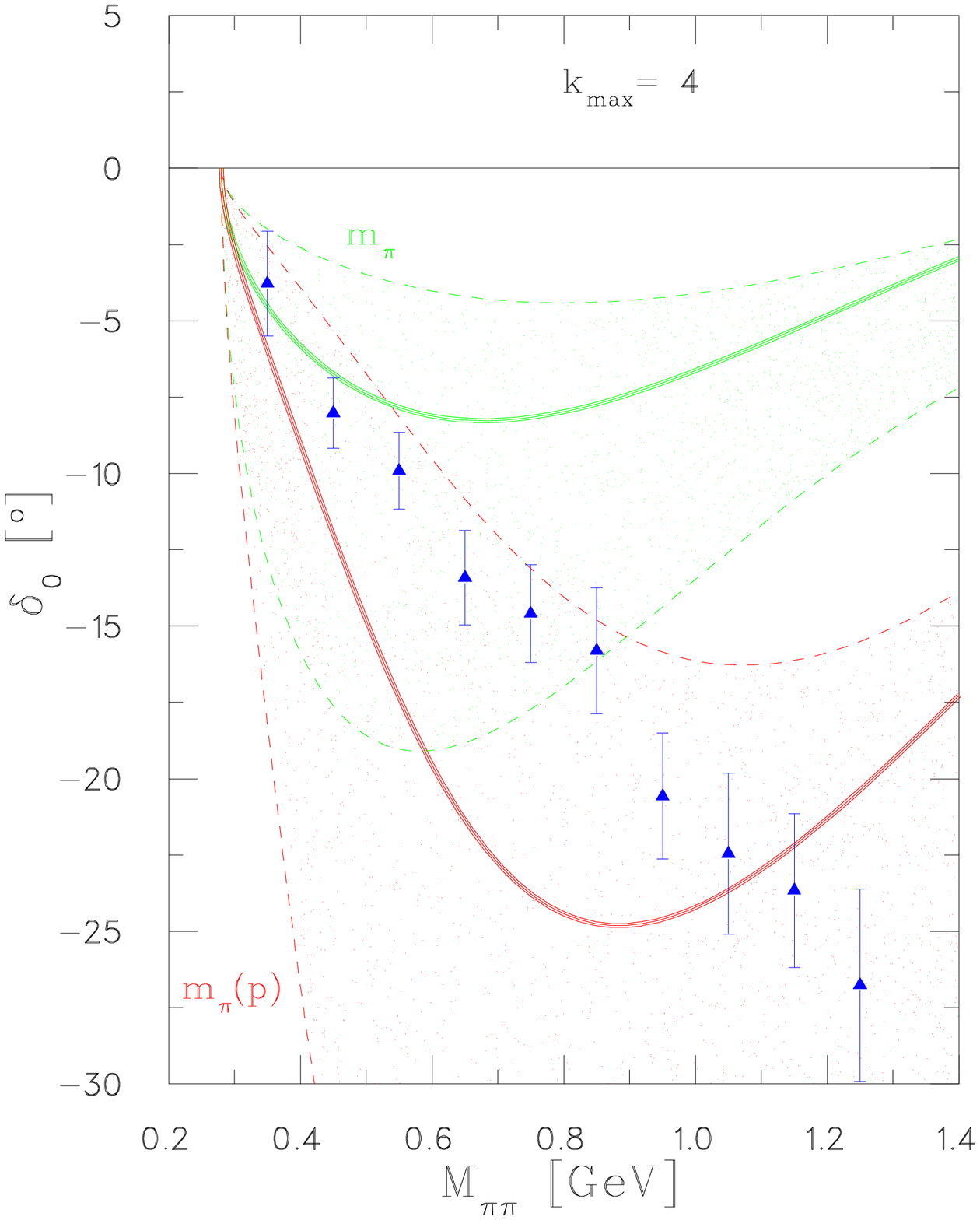}}
\end{center}
\caption{Scattering phase shifts $\delta^{I=2}_{\ell=0}$ calculated from the
$\pi$--$\pi$ potential of the lattice QCD simulation (thick lines).
Influence of momentum cut-off $k_{\max}=3$ (left) and $k_{\max}=4$ (right) is shown.
Results using the classical and relativistic dispersion relation are distinguished
by $m_\pi$ and $m_\pi(p)$, respectively.
Errors are represented by the dotted regions. Their boundaries (dashed lines) 
correspond to the phase shifts calculated with extremal (bootstrap)
potentials of Fig.~\protect\ref{figVe}. 
The experimental data are from \protect\cite{Hoo77}.}
\label{figD}\end{figure}

It is obvious that relativistic effects in the $\pi$--$\pi$
system are very large in the experimentally relevant kinematic region.
They are also inherent in the lattice simulation.
Nevertheless it is interesting to calculate scattering phase
shifts as they arise from the computed potential.  
This excercise is much in the spirit of section~\ref{secRGM}, where
it would be based on solving (\ref{RGM}) for scattering states.
We have used a Volterra integral equation of standard potential scattering
theory \cite{Tay72} with $V^{(\alpha)}(r)$ and obtained the scattering
phases from Jost functions. The results are shown in Fig.~\ref{figD},
where the two frames distinguish the truncations $k_{\max}=3,4$
respectively. The two upper solid curves correspond to the
physical reduced mass $m_\pi/2$ of the $\pi$--$\pi$ system, with
$m_\pi=0.28a^{-1}=0.14\mbox{GeV}$.
In order to estimate the magnitude of
the error due to the non-relativistic potential scattering theory
we have also used the relativistic dispersion relation
$m_\pi(p)=\sqrt{m_\pi^2+p^2}$.
The two lower solid curves in Fig.~\ref{figD} show the results.

Estimation of the statistical errors on the phase shifts is not a
straightforward task and requires some discretion. One option is to
select pairs of extremal potentials from
the bootstrap analysis of Fig.~\ref{figVe} and compute phase shifts for
those. This was done separately for $k_{\max}=3$ and $4$.
We have chosen the potentials which stand out in Fig.~\ref{figVe} as
the curves with largest and smallest values, respectively, at around
$r=0.6\mbox{fm}$ for $k_{\max}=3$ and around $r=0.8\mbox{fm}$
for $k_{\max}=4$. The resulting phase shifts are shown in Fig.~\ref{figD}
as the boundaries of the dotted regions.
Not surprisingly, as discussed above, the error spread is much larger for
$k_{\max}=4$. Results for both truncations are, nevertheless, consistent.

In the Volterra equation the parameterized potentials $V^{(\alpha)}(r)$
were used for the entire interval $r\in[0,\infty)$. Their deep attractive 
trough within $0\leq r\lesssim a$ is probed by large momenta
and is responsible for the scattering phase shifts turning upward beyond 
$M_{\pi\pi}>0.8-1.0\mbox{GeV}$.
This region is above the lattice cut-off.
For smaller $M_{\pi\pi}$ the intermediate and long distance region of the
potential, essentially as displayed in Fig.~\ref{figV}, is probed. 
There, it appears that our results are in qualitative
agreement with experimental findings. The data points shown as triangles
in Fig.~\ref{figD}
are from an analysis of the CERN-Munich $\pi^+p\rightarrow \pi^+\pi^+n$
experiment by Hoogland et al. \cite{Hoo77}.

\section{Conclusions and Outlook}

During the last two decades QCD has emerged as the underlying theory of
strong interactions. It is an important problem to put QCD to the test
of explaining the interactions between hadrons. The nonperturbative nature
of QCD in the energy domain of nuclear physics makes lattice field theory the
predestined technique to deal with this question.

We have made an effort
in this work to develop and apply lattice QCD techniques to the problem
of hadron-hadron interaction.
Within the example of a meson-meson system
we have outlined a strategy usable to extract a residual interaction
from a lattice simulation, which also may point a way to application
to other hadronic systems. 

On the practical side we have done a simulation
of the $\pi^+$--$\pi^+$ system and extracted a local potential for the
S-wave interaction. A coarse and large-volume lattice to make space for 
two hadrons, the extraction of small shifts of energy levels,
and steeply declining time correlation functions indigenous to two-hadron 
states all conspire to make the numerical work rather challenging. 
Nevertheless, our results indicate a repulsive interaction at
long and intermediate ranges, down to $r\approx 0.4\mbox{fm}$,
monotonously rising to $\approx 0.4\mbox{GeV}$ at that distance.
The corresponding scattering phase shifts are subject to large relativistic
corrections in the kinematical region of interest, however, within those
limitations, they compare favorably with experimental results.
The parameterization of our lattice results, matching a Yukawa
potential in the asymptotic region, allows us to make contact with
traditional boson-exchange models. In this picture we find that the
mass of an exchanged particle and its vertex coupling $g$ are
in the region of 1GeV
and 10, respectively, which are both reasonable for those quantities.

It appears that lattice techniques as tried in the present work
are feasible for tackling hadron-hadron interactions. Subsequent work
should make use of anisotropic lattices to ease the problems related
to steeply dropping correlation functions. It would also be highly
desirable to go beyond the `potential' picture and be able to extract
the scattering amplitude (t-matrix) in a more direct fashion numerically
from lattice QCD.
\vspace{2ex}

{\bf Acknowledgement:}
This work was supported by NSF PHY-9700502, by OTKA T023844
and by FWF P10468-PHY. One of the authors (HRF) is grateful for visiting
opportunities at the Institute for Nuclear Physics of the Technical
University of Vienna and at Thomas Jefferson Laboratory,
where significant advances were made. We thank Nathan Isgur for
pointing out \cite{Hoo77} to us. 
The final stages of the work were completed during a most inspiring stay
at the Special Research Center for the Structure of Subatomic Matter,
CSSM, where the manuscript was written. We would like to thank
A.W. Williams for useful discussions on some aspects of scattering
related to the lattice.

\newpage

\appendix

\section{Random Sources}~\label{appRanSrc}

We here give details of the random source technique used to compute
quark propagator matrix elements and of the smearing procedure.

Consider the linear equation
\begin{equation}
\sum_{\vec{y}y_4} \sum_{B\nu}\,G^{-1(f)}_{A\mu,B\nu}(\vec{x}x_4,\vec{y}y_4)\,
X^{(f;\,A'\mu'r\,x_4')}_{B\nu}(\vec{y}y_4) =
\delta_{AA'}\,\delta_{\mu\mu'}\,R^{(A'\mu'r\,x_4')}(\vec{x})\,
\delta_{x_4 x_4'} \,,
\label{DXR}\end{equation}
where $G^{-1(f)}$ is the (known) fermion matrix for flavor $f$, and
$R$ are complex Gaussian random vectors of length $L^3$ that live
on the space sites $\vec{x}$ of the lattice.
The meaning of the indices are $A,B\ldots=1,2,3$ {\em color}, 
$\mu,\nu\ldots=1,2,3,4$ {\em Dirac}, $\vec{x},\vec{y}\ldots$ {\em space}
($d=3$), $x_4,y_4$ {\em time}, and $r=1\dots N_R$ labels the random sources
$R$ for each source point. A prime~$'$ denotes a source point.
There is some freedom in choosing the latter.
In (\ref{DXR}) the sources are nonzero on one time slice $x_4'=t_0$ only.
A new source is chosen for each color, Dirac and time index.
The same set of sources is used for different flavors (meaning $\kappa$)
in order to take advantage of a multiple-mass solver \cite{Gla96}.

A version of a random source technique, called `maximal variance reduction'
\cite{Mic98} is currently in use for light-quark propagators in heavy-light
systems \cite{Mic99}. There, sources are spread across the entire lattice,
including all time slices. Since the variance of the source typically
is of order one, exponentially decaying time correlation functions will
quickly be engulfed in noise. To alleviate this problem a
subdivision of the lattice into disjoint regions, say $0 \leq t < T/2$ and
$T/2 \leq t < T$, is made and noise reduction is achieved for propagators
connecting those regions, see \cite{Mic98}.
On the other hand random sources $\propto \delta_{t t_0}$, as in (\ref{DXR}),
avoid the above problem from the outset for lack of noise at the sink.
They have proven sufficient in \cite{Can97} so we continue to employ them here.

The random sources $R$ are normalized according to
\begin{equation}
\sum_{\langle r\rangle} R^{(A'\mu'r\,x_4')}(\vec{x})
R^{(B'\nu'r\,y_4')\ast}(\vec{y}\,) = \delta_{A'B'}\,\delta_{\mu'\nu'}\,
\delta_{\vec{x}\vec{y}}\,\delta_{x_4' y_4'} \,.
\end{equation}
Here $\sum_{\langle r\rangle}$ denotes 
the random-source average, which we approximate numerically as
\begin{equation}
\sum_{\langle r\rangle}\ldots \simeq \frac{1}{N_R}\sum_{r=1}^{N_R}\ldots \,.
\end{equation}
Employing the solution vectors $X$ of (\ref{DXR}) an estimator for the
propagator matrix elements then is
\begin{equation}
G^{(f)}_{B\nu,A\mu}(\vec{y}y_4,\vec{x}x_4) =
\sum_{\langle r\rangle} X^{(f;\,A\mu\,r\,x_4)}_{B\nu}(\vec{y}y_4)
R^{(A\mu\,r\,x_4)\ast}(\vec{x}) \,.
\label{GXR}\end{equation}

Operator smearing \cite{Ale94} is defined through
\begin{equation}
\psi^{\{ 0\}}_{A}(\vec{x}t) = \psi_{A}(x) \quad\quad
\psi^{\{ s\}}_{A}(\vec{x}t) = \sum_{B} \sum_{\vec{y}}
K_{AB}(\vec{x},\vec{y}\,)\;\psi^{\{ s-1\}}_{B}(\vec{y}t) \,,
\label{Kpsi}\end{equation}
with $s\in{\Bbb N}$, and the matrix
\begin{equation}
K_{AB}(\vec{x},\vec{y}\,) = \delta_{AB}\,\delta_{\vec{x},\vec{y}}
+\alpha \sum_{m=1}^{3}
\left[ U_{m,AB}(\vec{x}\,t)\delta_{\vec{x},\vec{y}-\hat{m}}
+U^\dagger_{m,AB}(\vec{y}\,t)\delta_{\vec{x},\vec{y}+\hat{m}} \right] \,.
\label{smear2}\end{equation}
The real number $\alpha$ and the maximum value $S$ for $s=0\ldots S$ are parameters. 
We have used fuzzy link variables $U\in SU(3)$ in (\ref{smear2}).
Due to the linearity of (\ref{DXR}) the above iterative prescription
translates directly to the random source and solution vectors,
\begin{eqnarray}
R^{\{ 0\}(B'\nu''\,r\,t')}_{C'}(\vec{z}\,') &=&
\delta_{C'B'} R^{(B'\nu''\,r\,t')}(\vec{z}\,')
\quad\quad\mbox{(no sum over $B'$)} \\
R^{\{ s\}(B'\nu''\,r\,t')}_{C'}(\vec{z}\,') &=&
\sum_{\vec{y}\,'}  \sum_{B''}  K_{C'B''}(\vec{z}\,',\vec{y}\,')
R^{\{ s-1\}(B'\nu''\,r\,t')}_{B''}(\vec{y}\,')
\end{eqnarray}
\begin{eqnarray}
X^{\{ 0\}(f;\,B'\nu''\,r\,t')}_{C\mu''}(\vec{z}t) &=&
X^{(f;\,B'\nu''\,r\,t')}_{C\mu''}(\vec{z}t) \\
X^{\{ s\}(f;\,B'\nu''\,r\,t')}_{C\mu''}(\vec{z}t) &=&
\sum_{\vec{x}} \sum_{A} K_{CA}(\vec{z},\vec{x})
X^{\{ s-1\}(f;\,B'\nu''\,r\,t')}_{A\mu''}(\vec{x}t) \,.
\end{eqnarray}
Finally, replacing $R\rightarrow R^{\{ S\}}$ and  $X\rightarrow X^{\{ S\}}$
in (\ref{GXR}) yields the propagator $G^{\{ S\}}$ for smeared fermion
fields as used in (\ref{contrac}) and (\ref{Gflavor}).

\section{Elementary Bose Field}~\label{appElem}

Let $\hat{\phi}(x)$ be an elementary Bose field defined on the sites 
$x=(\vec{x},t)$ of the lattice. It is understood that $\hat{\phi}$ is subject 
to canonical quantization. Let 
$\hat{\cal L}=\hat{\cal L}_0+\hat{\cal L}_I$ be a Lagrangian such that
$\hat{\cal L}_0(\hat{\phi},\partial\hat{\phi})$ is the free part and 
$\hat{\cal L}_I=\hat{\cal L}_I(\hat{\phi})$ is a (small) interaction.
Thus $\hat{\cal L}$ gives rise to a Hamiltonian $\hat{H}=\hat{H}_0+\hat{H}_I$
with according interpretation of its terms.
In analogy to (\ref{onephi}) and (\ref{twophi}) further define
$
\hat{\phi}_{\vec{p}}(t)=
L^{-3}\sum_{\vec{x}}\,e^{i\vec{p}\cdot\vec{x}}\hat{\phi}(\vec{x},t)
$
and
$
\hat{\Phi}_{\vec{p}}(t)=\hat{\phi}_{-\vec{p}}(t)\,\hat{\phi}_{+\vec{p}}(t) \,.
$
We have in mind a perturbative calculation of the
correlation matrix \cite{Can97} 
\begin{equation} 
\hat{C}^{(4)}_{\vec{p}\,\vec{q}}(t,t_0)=
\langle 0|\hat{\Phi}^{\dagger}_{\vec{p}}(t)\,
\hat{\Phi}_{\vec{q}}(t_0)|0\rangle
\label{eqA4}\end{equation}
assuming a nondegenerate vacuum state $|0\rangle$ with $\hat{H}_0|0\rangle=0$.
Standard perturbation theory gives rise to the time evolution operator
\begin{equation} 
\hat{U}(t,t_0)=\sum_{N=0}^{\infty}\,\frac{(-1)^N}{N!}\,
\int_{t_0}^t\!dt_1\ldots\int_{t_0}^t\!dt_N\,
\mbox{T}[\hat{H}_I(t_1)\ldots\hat{H}_I(t_N)] \,,
\label{eqA8}\end{equation}
where $\hat{H}_I(t)=e^{\hat{H}_0(t-t_0)}\,\hat{H}_I\,e^{-\hat{H}_0(t-t_0)}$
refers to the interaction picture. Working out (\ref{eqA4}) with (\ref{eqA8})
induces a perturbative expansion of the correlator
\begin{equation}
\hat{C}^{(4)}_{\vec{p}\,\vec{q}}(t,t_0)=
\sum_{N=0}^{\infty}\,\hat{C}^{(4;N)}_{\vec{p}\,\vec{q}}(t,t_0) \,.
\label{eqA10}\end{equation}

Explicit forms of the $N=0$ and the $N=1$ terms are conveniently
expressed in terms of wave functions
$\psi^{(0)}_{n\nu}(\vec{p}\,)$
defined through
\begin{equation} 
c^{(0)}_{n\nu}\psi^{(0)}_{n\nu}(\vec{p}\,) =
\langle n\nu|\hat{\Phi}_{\vec{p}}(t_0)|0\rangle^{\ast} \,,
\label{eqA12}\end{equation}
where 
$|n\nu\rangle$
is a complete set of eigenstates of $\hat{H}_0$ with
$\hat{H}_0 |n\nu\rangle = W^{(0)}_n |n\nu\rangle$, and
$c^{(0)}_{n\nu}$
are appropriate normalization factors chosen such that the
$\psi^{(0)}_{n\nu}$
are orthonormal.
The corresponding correlator terms are
\begin{eqnarray} 
\hat{C}^{(4;N=0)}_{\vec{p}\,\vec{q}}(t,t_0) &=&
\sum_{n\nu} |c^{(0)}_{n\nu}|^2 e^{-W^{(0)}_n(t-t_0)} 
\psi^{(0)}_{n\nu}(\vec{p}\,) \psi^{(0)\ast}_{n\nu}(\vec{q}\,)
\label{eqA13} \\
\hat{C}^{(4;N=1)}_{\vec{p}\,\vec{q}}(t,t_0) &=& 
-\sum_{n\nu}\,\sum_{m\mu}\,\psi^{(0)}_{n\nu}(\vec{p}\,) 
\psi^{(0)\ast}_{m\mu}(\vec{q}\,) \nonumber \\ & &
\langle n\nu|\hat{H}_I|m\mu\rangle c^{(0)\ast}_{n\nu} c^{(0)}_{m\mu}
\exp\left[-\frac{W^{(0)}_n+W^{(0)}_m}{2}(t-t_0)\right] \nonumber \\ & &
\left\{ (t-t_0)\delta_{nm} + 
\frac{\sinh\left[\frac{W^{(0)}_n-W^{(0)}_m}{2}(t-t_0)\right]}
{\frac{W^{(0)}_n-W^{(0)}_m}{2}}(1-\delta_{nm})
\right\} \,.
\label{eqA17}\end{eqnarray}
Without loss of generality the normalization constants $c^{(0)}_{n\nu}$ may be 
chosen real and positive.
We now observe that the two normalization factors and the exponential 
in (\ref{eqA17}) may be removed by multiplying 
$\hat{C}^{(4;N=1)}$ from both sides with the inverse square root of
$\hat{C}^{(4;N=0)}$. Hence the matrix elements of
\begin{equation} 
\hat{\cal C}^{(4;N=1)}(t,t_0) = {\hat{C}^{(4;N=0)}(t,t_0)}^{-1/2}\,
\hat{C}^{(4;N=1)}(t,t_0)\,{\hat{C}^{(4;N=0)}(t,t_0)}^{-1/2} 
\label{eqA18}\end{equation}
in the basis $\psi^{(0)}_{n\nu}(\vec{p}\,)$ are products of 
$\langle n\nu|\hat{H}_I|m\mu\rangle$ and the expression inside 
$\left\{\ldots\right\}$ of (\ref{eqA17}).  The $t$ derivative of the latter is 
equal to one at $t=t_0$.  Thus we have
\begin{equation} 
\left[ \frac{\partial \hat{\cal C}^{(4;N=1)}_{\vec{p}\,\vec{q}}(t,t_0)}
{\partial t} \right]_{t=t_0} =
-\sum_{n\nu}\,\sum_{m\mu}\,\psi^{(0)}_{n\nu}(\vec{p}\,) 
\langle n\nu|\hat{H}_I|m\mu\rangle \psi^{(0)\ast}_{m\mu}(\vec{q}\,) =
\hat{H}_I \,,
\label{eqA19}\end{equation}
which holds independently of the basis.
Finally, we may replace $\hat{C}^{(4;N=1)}$ in (\ref{eqA18})-(\ref{eqA19}) 
with the full correlation matrix $\hat{C}^{(4)}$ since this will only
introduce an $N=2$ error.
This leads us to define an `effective correlator'
\begin{equation} 
\hat{\cal C}^{(4)}(t,t_0) = {\hat{C}^{(4;N=0)}(t,t_0)}^{-1/2}\,
\hat{C}^{(4)}(t,t_0)\,{\hat{C}^{(4;N=0)}(t,t_0)}^{-1/2} \,.
\label{eqA18a}\end{equation}

We have shown at this point that $\hat{\cal C}^{(4)}$ has the power series
expansion $\hat{\cal C}^{(4)}(t,t_0) = {\Bbb 1} +
\hat{H}_I(t-t_0) + {\cal O}(h^2)$ where the second-order remainder must
depend on the product $h=\hat{H}_I(t-t_0)$ for dimensional reasons.
In the next order $N=2$ a calculation similar to the above reveals that
\begin{equation} 
\hat{\cal C}^{(4)}(t,t_0) = {\Bbb 1}+\hat{H}_I(t-t_0)
+\frac{1}{2}\left(\hat{H}_I(t-t_0)\right)^2+{\cal O}(h^3) \,.
\label{eqA23}\end{equation}
Whether or not these are the initial terms of a converging power series
expansion for $\hat{\cal C}^{(4)}(t,t_0)$ will depend on the actual $H_I$.
In case the series converges `everywhere' the expansion will define a correlator
even for large $t-t_0$.  
Limiting ourselves to order $N=2$ perturbation theory (\ref{eqA23}) is the same as
\begin{equation} 
\hat{\cal C}^{(4)}(t,t_0) = e^{-\hat{H}_I(t-t_0)} \,.
\label{eqA22}\end{equation}

The utility of these results in the framework of a lattice simulation
lies in the 
analogy that can be drawn between $\hat{C}^{(4;N=0)}$ and the free
correlator $\overline{C}^{(4)}$, and between $\hat{C}^{(4)}$ and the full
correlator $C^{(4)}$. The analogue of (\ref{eqA22}) may then be considered
as a definition of an effective interaction.

\end{document}